\documentclass[prl,floatfix,reprint]{revtex4}
\usepackage{graphicx,amssymb,amsmath,color}

\usepackage{booktabs}

\begin{document}

\title{A stochastic model of randomly accelerated walkers for human mobility}

\author{Riccardo Gallotti}
\affiliation{Institut de Physique Th\'{e}orique, CEA-Saclay, Gif-sur-Yvette, France}
\author{Armando Bazzani}
\affiliation{Department of Physics and Astronomy, University of Bologna, INFN Bologna section}
\author{Sandro Rambaldi}
\affiliation{Department of Physics and Astronomy, University of Bologna, INFN Bologna section}
\author{Marc Barthelemy\footnote{Correspondance to: marc.barthelemy@cea.fr}}
\affiliation{Institut de Physique Th\'{e}orique, CEA-Saclay, Gif-sur-Yvette, France}
\affiliation{CAMS (CNRS/EHESS) 190-198, avenue de France, 75244 Paris Cedex 13, France}

\begin{abstract}
The recent availability of large databases allows to study macroscopic properties of many complex systems. However, inferring a model from a fit of empirical data without any knowledge of the dynamics might lead to erroneous interpretations~\cite{Benhamou:2007}. We illustrate this in the case of human mobility~\cite{Brockmann:2006,Gonzalez:2008,Rhee:2011} and foraging human patterns~\cite{Raichlen:2013} where empirical long-tailed distributions of jump sizes have been associated to scale-free super-diffusive random walks called L\'evy flights~\cite{Book:foraging}. Here, we introduce a new class of accelerated random walks where the velocity changes due to acceleration kicks at random times, which combined with a peaked distribution of travel times~\cite{Gallotti:2015ttb}, displays a jump length distribution that could easily be misinterpreted as a truncated power law, but that is not governed by large fluctuations. This stochastic model allows us to explain empirical observations about the movements of 780,000 private vehicles in Italy, and more generally, to get a deeper quantitative understanding of human mobility.
\end{abstract}

\maketitle

With the development of Information and Communication Technologies~\cite{Vespignani:2012}, the investigations' focus shifted from the traditional travel diary surveys~\cite{Axhausen:2002,Yan:2013,Liang:2013} to several new datasources. In particular, it became possible to follow individual trajectories from mobile phone calls~\cite{Gonzalez:2008,Song:2010,Kang:2012}, location-sharing services~\cite{Cheng:2011,Noulas:2012,Liu:2014} and microblogging~\cite{Hawelka:2014}, or directly extracted from public transport ticketing system~\cite{Roth:2011,Liang:2013} or GPS tracks of taxis~\cite{Liang:2012,Liu:2012,Liang:2013,Wang:2015,Liu:2015,Tang:2015}, private cars~\cite{Bazzani:2010,Gallotti:2012} or single individuals~\cite{Rhee:2011,Zhao:2015}. For most data sources, the spatial positions $r$ are the most reliable quantity. This information can be used for studying two different aspects of human mobility: (i) how far and (ii) where we are moving. The second question is far more complex than the first and can be approached with several different methodologies, from aggregated origin-destination matrices~\cite{Sagarra:2015} suitable for mobility prediction~\cite{Simini:2012} or land use analysis~\cite{Louail:2014}, to individual mobility networks and patterns suitable for describing the natural tendency to return frequently to a few locations (such as homes, offices, etc)~\cite{Garling:2003,Gonzalez:2008,Song:2010,Bazzani:2010,Gallotti:2012,Gallotti:2013}. On the other hand, the first question, generally characterized by the distribution of the spatial displacements $P(\Delta r)$ across all users, although apparently simple is still far from being completely understood. Indeed, even if the study of $P(\Delta r)$ has become a trademark for recent studies on human mobility, there are still no consensus about the functional form of this distribution. At a large scale (national or inter-urban), one may observe a long tail behavior~\cite{Brockmann:2006,Gonzalez:2008,Song:2010,Cheng:2011,Liu:2012,Noulas:2012,Yan:2013,Hawelka:2014,Liu:2015} characterized by a power law decay for long displacements. At a smaller (urban) scale, the distribution seems to have an exponential tail \cite{Bazzani:2010, Gallotti:2012,Liang:2012,Kang:2012,Liang:2013,Wang:2015,Liu:2015}. The unclear nature of this probability distribution makes its interpretation difficult and dependent on the dataset used, the scale and possible empirical errors (see Table~\ref{table1}). It is therefore necessary to to obtain data as clean as possible, to propose a model that can be tested against empirical results. So far power law fits were used and led the authors to draw conclusions about the nature and mechanisms of the mobility, but this way of proceeding could actually lead to erroneous conclusions.

Human travelling behaviour can be described as a sequence of resting times of duration $\tau$ and instantaneous jumps $\Delta r$ in space~\cite{Song:2010}. These two processes need to be separated for modelling human mobility, since costs are in general associated to trips while a positive utility can be associated to activities performed during stops~\cite{Axhausen:1992}. This approach is therefore fundamental for a large variety of complex processes in cities where mobility is central, such as traffic forecasting~\cite{Axhausen:1992}, epidemics spreading~\cite{Colizza:2007,Balcan:2009} or the evolution of cities~\cite{Makse:1995,Bettencourt:2013,Louf:2014}.

The distribution of displacements $P(\Delta r)$ plays a central role and several studies suggested the existence a fat tails (power law distribuitons) with a wide range of values for the exponent (see Table~\ref{table1}). In contrast, at the urban scale, displacements seem to be described by exponential tails~\cite{Bazzani:2010,Gallotti:2012,Liang:2012}. A similar, unresolved controversy also exists about animal's foraging movements: relying only on a fit of the empirical data, the same distribution can be understood in different ways, leading to contrasting conclusions on the nature of the underlying process~\cite{Edwards:2007,Edwards:2011,Jansen:2012}. Remarkably enough, what appears to be underevaluated in the study of human mobility is the relevance of travel itself. The proposed models usually neglect the role of travel time and the moving velocity, also those indicating the hierarchy of transportation networks as the cause of this long tail~\cite{Han:2011}, assuming instantaneous jumps. This is essentially a consequence of the limitations inherent to datasources. Phone calls or social networks capture the spatial character of individuals' movements~\cite{Lenormand:2014}, but they are limited by sampling rates or by the bursty nature of human communications~\cite{Barabasi:2005} and are thus not suitable for an exhaustive temporal description of human mobility. 

\begin{table*}[ht!]\centering
\begin{tabular*}{\textwidth}{@{\extracolsep{\fill}}lcccccccccc@{}}
\hline
& Data Source & Trajectories & $\beta$ & $\kappa$ & $\Delta r_0$\\
\hline
& Dollar Bills~\cite{Brockmann:2006}		& 464K	& 1.59		& $\infty$		& 0\\
& Mobile Phones~\cite{Gonzalez:2008}	& 100K	& 1.75		& 400 km		& 1.5 km \\
& Mobile Phones~\cite{Gonzalez:2008}	& 206	& 1.75		& 80 km		& 1.5 km \\
& Mobile Phones~\cite{Song:2010}		& 3M 	& 1.55		& 100 km		& 0\\
& Location Sharing~\cite{Cheng:2011}	& 220K	& 1.88		& $\infty$		& 0\\
& GPS tracks~\cite{Rhee:2011}			& 101	& [1.16,1.82]	& $\infty$		& 0\\
& Location Sharing~\cite{Noulas:2012} 	& 900K 	& 1.50		& $\infty$		& 2.87 km\\
& Location Sharing~\cite{Noulas:2012} 	& 900K 	& 4.67		& $\infty$		& 18.42 km\\
& Taxis~\cite{Liang:2012}				& 12K	& 0			& 4.29 km		& -\\
& Taxis~\cite{Liu:2012}				& 7K		& 1.2			& 10 km		& 0.31 km\\
& Mobile Phones~\cite{Kang:2012}		& 7K		& 0			& [2, 5.8] km 	& -\\
& Travel Diaries~\cite{Yan:2013}		& 230	& 1.05 		& 50 km		& 0\\
& Tweets~\cite{Hawelka:2014} 			& 13M 	& 1.62		& $\infty$		& 0\\
& Location Sharing~\cite{Liu:2014} 		& 521K	& 0			& 300 km		& -\\
& Taxis~\cite{Wang:2015}				& 30K	& 0			& [2, 4.6] km 	& -\\
& Taxis~\cite{Tang:2015}				& 1100	& [0.50,1.17]	& [4.5, 6.5] km	& 0\\
\hline
\end{tabular*}
\caption{Parameter values for the fit of the displacement distribution with a truncated power law found in previous studies. This list includes studies on different datasources and spatial or temporal scales. Only fits consistent with the function $P(\Delta r) = (\Delta r+\Delta r_0)^{-\beta}\exp(-\frac{\Delta r}{\kappa})$ are presented here. The case $\kappa = \infty$ is associated to non-truncated power laws, while $\beta = 0$ to exponential distributions. $\Delta r_0 = 0$ correspond to cases where this parameter was omitted in the fit, while $\beta=0$ when this value is not defined.  Further studies propose: i) a polynomial form close to an exponential behavior for Private Cars~\cite{Bazzani:2010}; ii) two different behaviours for urban and inter-urban trajectories for Cars and Taxis~\cite{Gallotti:2012,Liu:2015}; iii) a lognormal distribution for individual GPS tracks~\cite{Zhao:2015}.}
\label{table1}
\end{table*}

In this paper, we show that the observed truncated power laws in the jump size distribution can be the consequence of simple processes such as random walks with random velocities~\cite{Zaburdaev:2008}. We test this model over a large GPS database describing the mobility of $780,000$ private vehicles in Italy, where travels and pauses can be easily separated, as the transition is identified by the moment when the engine is turned on or off (but we introduce a lower threshold of 5 minutes in the elapsed time, to distinguish real stops from accidentaly switched off of the engine during a trip). This allows us to evaluate accurately not only the displacements $\Delta r$, but also travel-times $t$, speeds $v$, and rest times $\tau$. 

\clearpage

\subsection*{Travel-times}

\begin{figure*} [ht!]
\begin{center}
\begin{tabular}{c}
\includegraphics[angle=0, width=0.45\textwidth]{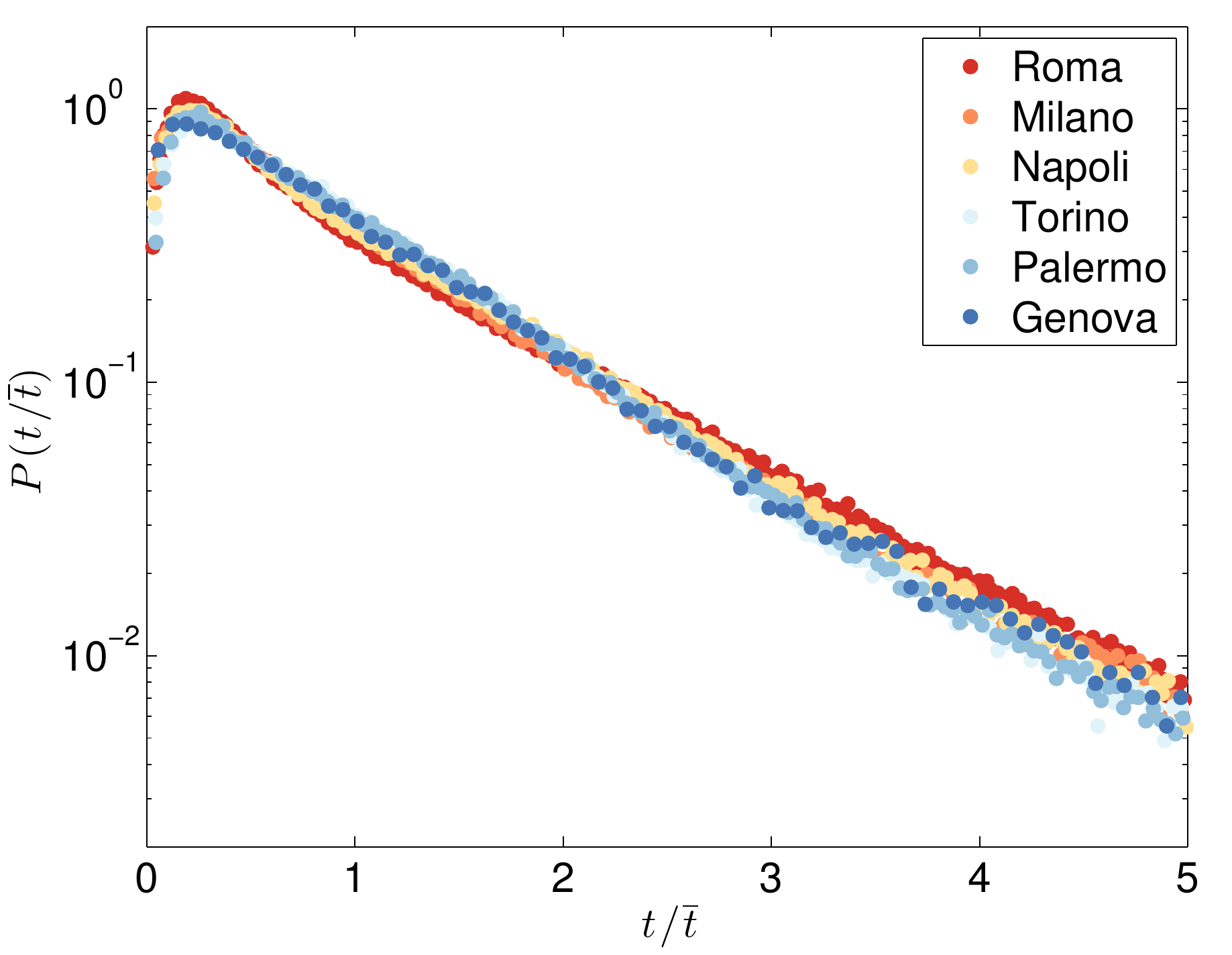}
\end{tabular}
\end{center}
\caption{{\bf Travel time in the 6 largest Italian cities.} The probability distribution is an exponential function and the difference between cities is therefore fully encoded in a single parameter: the average travel-time $\overline t$.
}
\label{figTimes}
\end{figure*}

We first analyze the distribution $P(t)$ of the travel times of individual vehicle in different cities. We observe in Fig.~\ref{figTimes} that $P(t)$ is 
characterized by an exponential decay as in~\cite{Gallotti:2015ttb,Zhao:2015}. For public transport systems, we also observe a rapidly decreasing tail for the travel-times between metro stations in London evaluated from the {\it Oyster card} ticketing system (see~\cite{Roth:2011} and Fig.~\ref{figS1} left).  A short tail is expected for these distributions, since the total daily travel-time spent in public~\cite{Kolbl:2003} or private~\cite{Gallotti:2015ttb} transportation has an exponential tail.

The individual mobility is composed by essentially two different phases, travels and rests, both having a finite duration. Studying times instead of distances allows to exclude the variability due to the different speeds, which may contribute to the perceived difference in travel costs. When travel costs are homogeneous, their distribution is expected to be exponential~\cite{Yan:2013}. Similarly to what is observed for total daily travel-times~\cite{Kolbl:2003,Gallotti:2012,Gallotti:2015ttb}, the exponential distribution appears in a variety of context such as the difference between independent events or maximizing the entropy at fixed average. We observe that the distribution of cars' travel-times is indeed characterized by an exponential tail (Fig.~\ref{figTimes}). 

For public transport systems, we also observe a rapidly decreasing tail for the travel-times between metro stations in London evaluated from the {\it oyster card} ticketing system. (See~\cite{Roth:2011} and Fig.~\ref{figS1} left). Travel times for cars however depend on the city population~\cite{Louf:2014}: the largest the city, and the worse congestion effects. This is confirmed by our results: if we bin the drivers according to their city of residence (see Methods), we find that the average travel-time in different cities falls in the range $[9~\mathrm{min},18~\mathrm{min}]$, with a growth correlated with population $\langle t \rangle \propto P^{\mu}$, where $\mu = 0.07\pm0.02$ (Fig.~\ref{figS2}). As a consequence of this, the average travel time differs from a city to another one (Fig.~\ref{figS1} right), but if we rescale the time by its average value, we observe in Fig.~\ref{figTimes} that  the distribution $P(t/\overline t)$ appears to be universal across different cities and given by
\begin{equation}
P(t)=\frac{1}{\overline t}\mathrm{e}^{-t/\overline t}
\label{eq:time}
\end{equation}
The average value $\overline t$ contains then all the information needed for describing cars' travel times in a particular city.

\subsection*{Rest times}

\begin{figure*}[ht!]
\begin{center}
\begin{tabular}{cc}
\includegraphics[angle=0,width=0.45\textwidth]{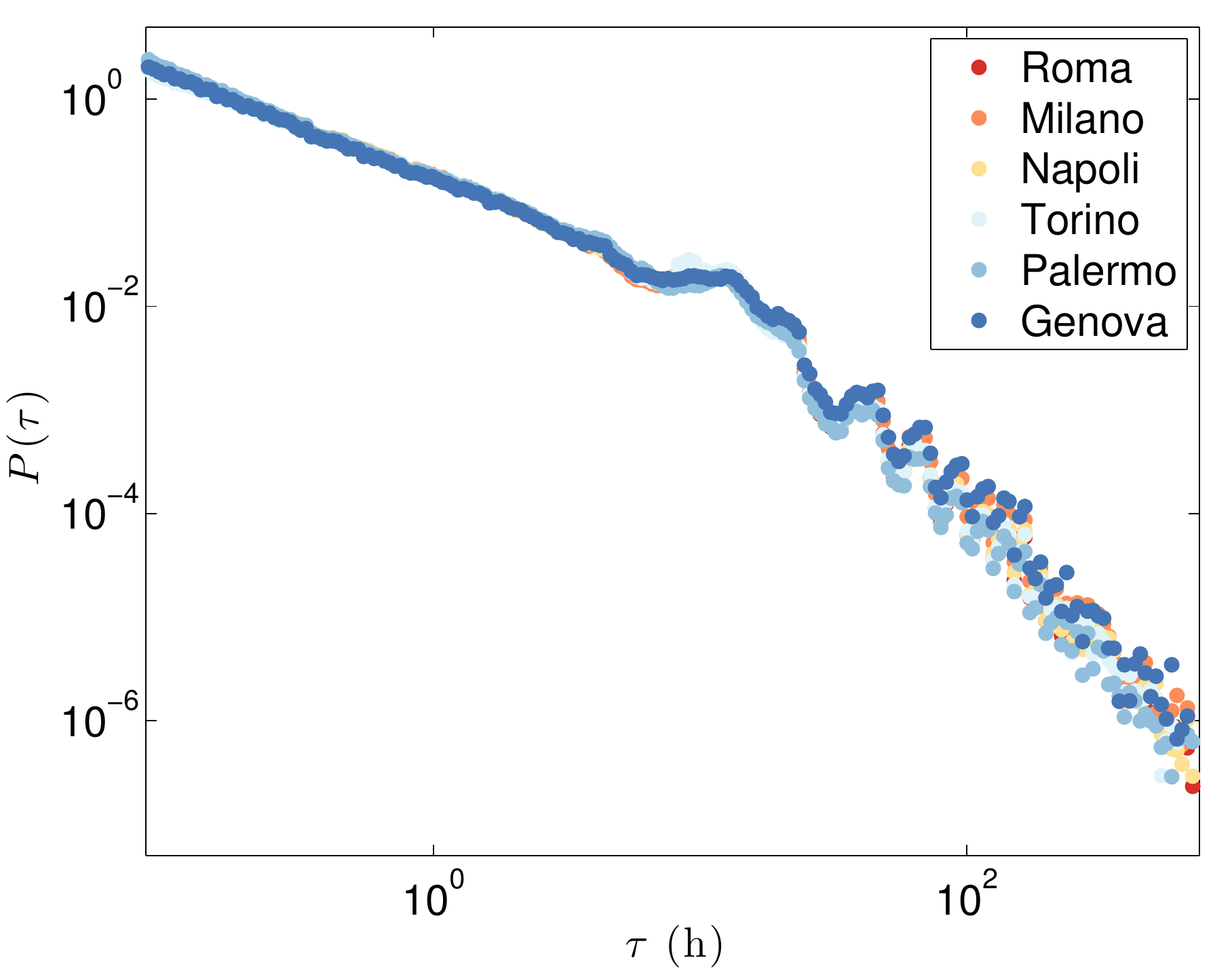}&
\includegraphics[angle=0,width=0.45\textwidth]{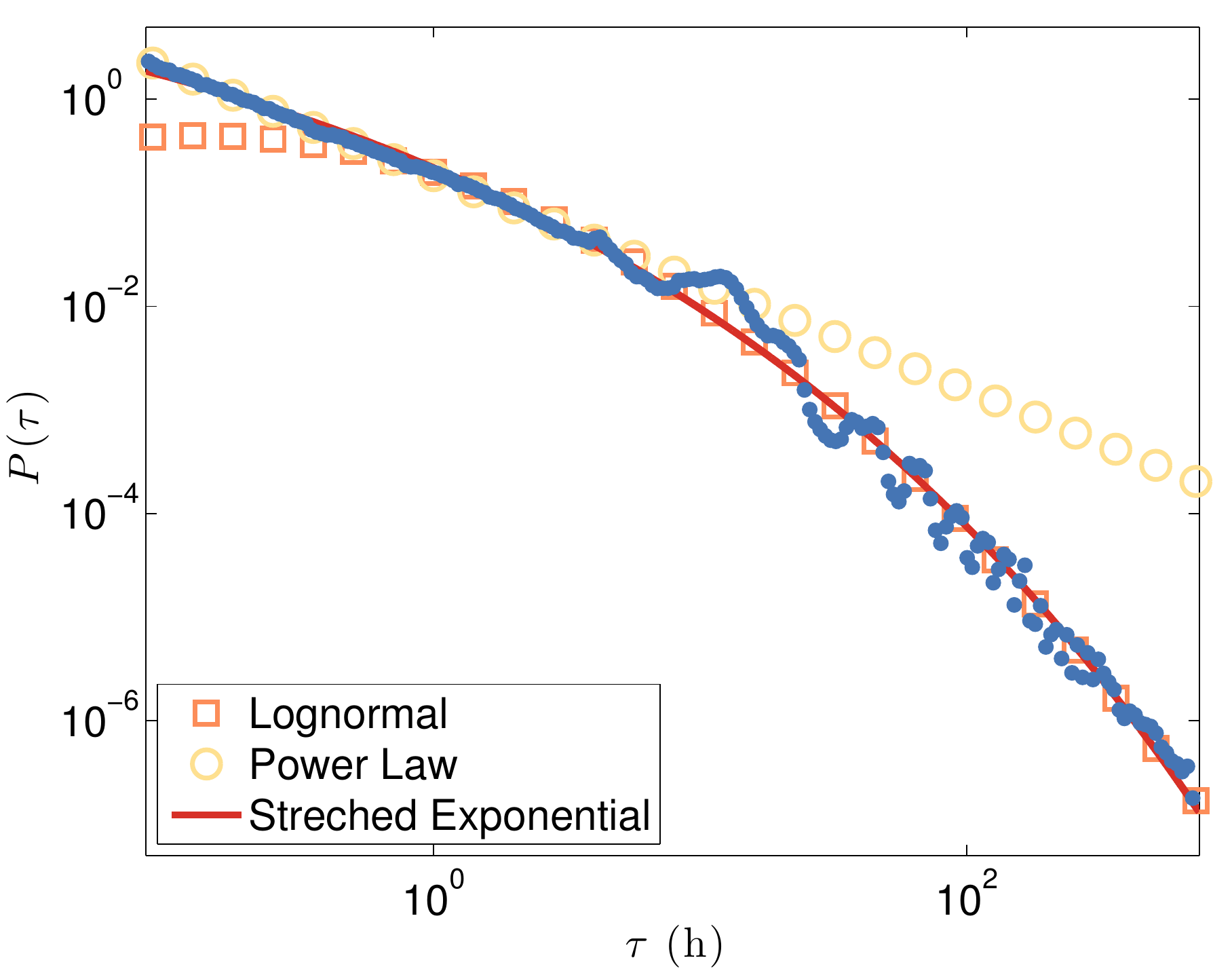}\\
\end{tabular}
\end{center}
\caption{{\bf Pause times distribution of private cars.} {\bf (Left)}
  The distribution shows little variation among the six largest cities
  in Italy. {\bf (Right)} Considering all the individuals across the
  whole Italy, we show that the distribution appear to follow a power law
  $P(\tau)\propto \tau^{-\eta}$ with $\eta=1.03\pm0.01$, valid for
  pauses shorter than 4 hours (yellow circles), which represents 74\%
  of the pauses. Conversely, a lognormal distribution
  $\ln\mathcal{N}(\mu,\,\sigma)$ with $\mu = 1.6\pm0.6$ h and
  $\sigma = 1.6\pm0.1$ h could be proposed to characterize the tail
  for pauses longer than 1 hour (orange squares). A stretched
  exponential: $P(\tau)\propto\exp(-(\tau/\tau_0)^\beta)$ with
  $\beta =0.19\pm0.01$ and $\tau_0=10\pm5\cdot10^{-5}$ h successfully
  fits the whole curve at all order of magnitude (red solid
  line). 
  }
\label{figS3}
\end{figure*}

The distribution of rest times $P(\tau)$ does not display remarkable differences between cities for stops within 24 hours (see Fig.~\ref{figS3} left). Consistently with the delay time distribution in email communications~\cite{Barabasi:2005,Proekt:2012}, this distribution is close to a power law with exponent $\simeq-1$~\cite{Gallotti:2012} (for times shorter than 12 hours). A lognormal fit for the tail seems also reasonable~\cite{Stouffer:2006}. In Fig.~\ref{figS3} right, we show the best fits with these functions, while we also propose a fit with a stretched exponential that, although it represent a good fit for all times between 5 minutes and 30 days, lacks any theoretical explanation.

\subsection*{Dependence of velocities over travel-time}

\begin{figure*}[ht!]
\begin{center}
\begin{tabular}{cc}
\includegraphics[angle=0,width=0.45\textwidth]{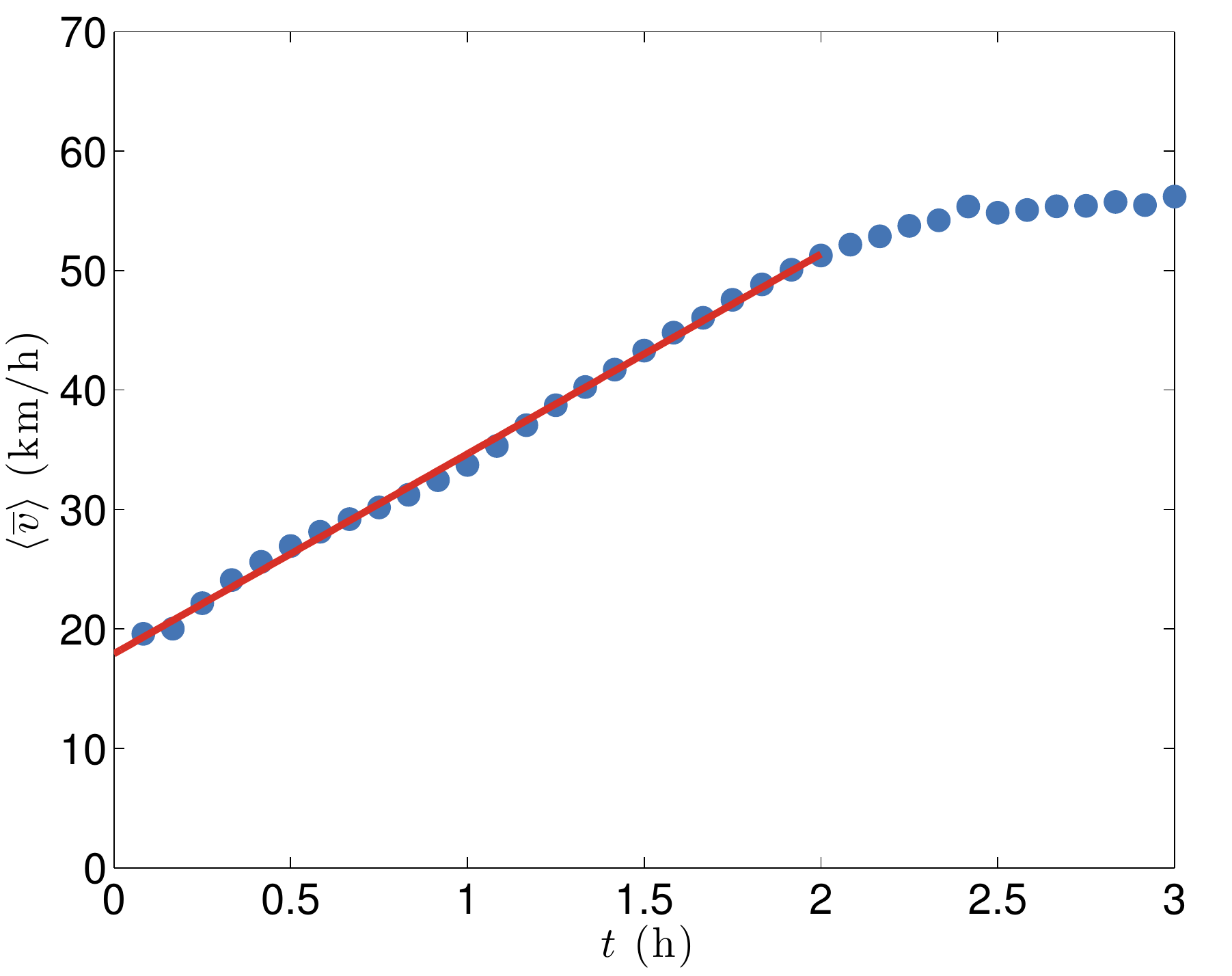}&
\end{tabular}
\end{center}
\caption{{\bf Acceleration of private transportation.} Empirical average velocities versus the duration of the trip for our GPS dataset. The solid line represents a constant acceleration $v = v_0 + at$ with $v_0 = 17.9$ km/h and $a = 16.7$ km/h$^2$.}
\label{speeds}
\end{figure*}

In order to describe the evolution of velocities, we consider a very general one-dimensional example of randomly accelerated walk, with uncorrelated accelerations
\begin{equation}
\dot v = \sqrt{D}\xi(\theta)
\label{eq:v}
\end{equation}
where $\xi$ is the white noise and $\theta$ is time. This process yields to a Gaussian-like distribution of velocity and which leads for its absolute value $P(|v|,\theta) \propto \exp\left(-\frac{|v|^2}{2D\theta}\right)$. Combining this result with the exponential distribution of travel-times $P(t)\sim \mathrm{e}^{-t/\overline t}$, and using a saddle-point approximation, we obtain for large displacements a stretched exponential of the form
\begin{equation}
P(\Delta r) \propto  \Delta r^{-\gamma} \exp(-C \Delta r^\delta)
\label{randomAccPr}
\end{equation}
with $\gamma = 1/4$ and $\delta =  1/2$.
We show in Fig.~\ref{distances} the best-fit with this curve, indicated as ``Stretched Exponential''. This fit with a single parameter ($C=0.63$ km$^{-0.5}$) would already offer a reasonable description of an empirical distribution that is also fitted by a truncated power law with $3$ parameters~\cite{Gonzalez:2008} (See also Fig.~\ref{rCities}). In addition, the simple random model for velocities described by Eq.~(\ref{eq:v}) induces a relation~\cite{Gallotti:2013phd} between the travel-time $t$ and the average speed $\langle |v| \rangle$ of the form $\langle |v| \rangle \propto t^{1/2}$, which as we will show, is not consistent with empirical data. Similarly to what was proposed for the foraging movements of animals and humans, we are mixing here short and slow with long and fast jumps, but instead of alternating between behavioral phases where the trajectories have different speed and autocorrelations~\cite{Codling:2011}, we have here a continuous distribution of velocities due to the structure of the mobility network.
However, for both private cars (Fig.~\ref{speeds}) and public transport (Fig.~\ref{publicTransport} left), we find that average velocities grow linearly for large travel times, and not as a square root. 
We show here that this apparent constant acceleration results from an optimal use for longer trips of the hierarchical nature of the transportation networks. Indeed, it is likely that faster transportation modes or faster roads are used more frequently for longer trajectories~\cite{Gallotti:2012,Gallotti:2014}. In the following, we will propose a simple stochastic phenomenon based on this idea.

\subsection*{Randomly accelerated walkers}
\begin{figure*}[ht!]
\begin{center}
\begin{tabular}{cc}
\includegraphics[angle=0,width=0.45\textwidth]{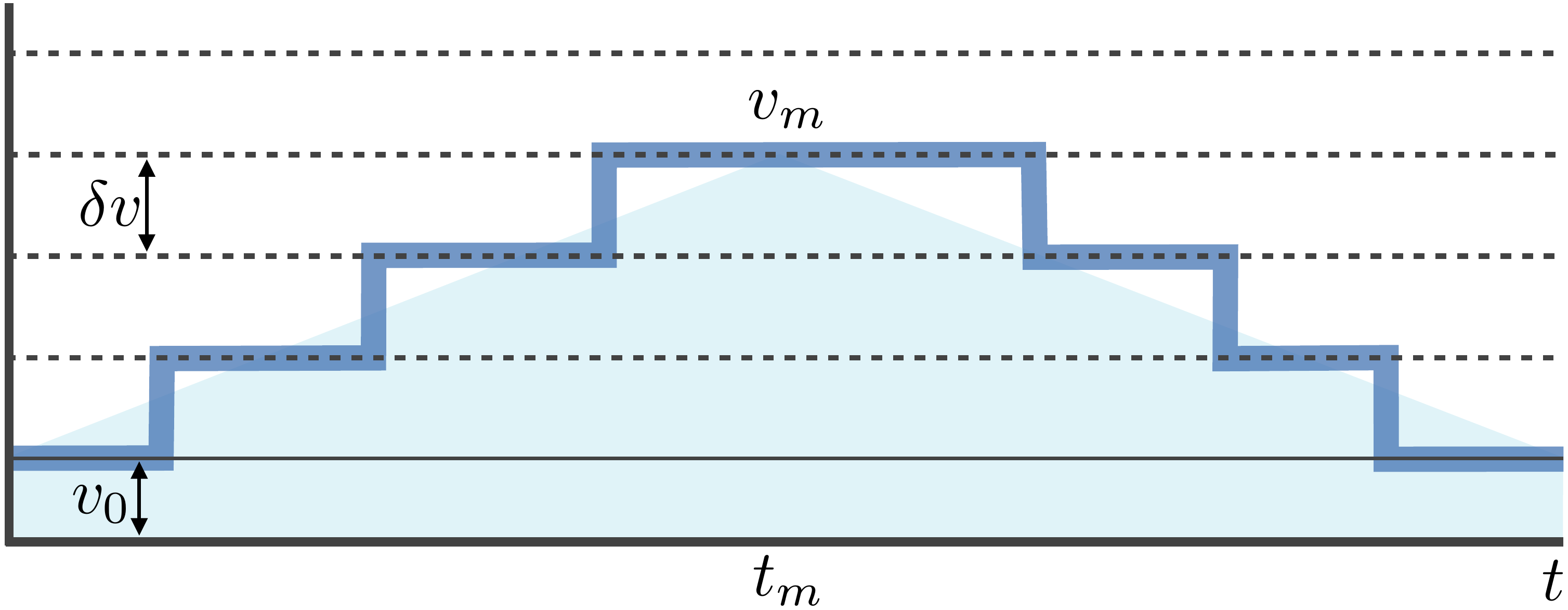}
\end{tabular}
\end{center}
\caption{{\bf Schematic representation of the model}. The trajectory starts with an acceleration phase, where the velocity is increased by constant kicks equal to $\delta v$. These kicks happen at random times with an uniform probability per unit time $p$. When approaching to the destination, there is a deceleration phase with kicks of -$\delta v$ and the same probability $p$. Given the symmetry of the problem, the maximal speed $v^*$ is reached on average at $t_m =  t/2$. This maximal speed depends on $\delta v$ and $p$: $v_m = v_0 + k(t_mp)\delta v$, where $k(\lambda)$ follows a Poisson distribution of average $\lambda$. The average speed $\overline v$ of the trajectory can be evaluated by approximating the step function which defines the light blue area, as $\overline v = (v_0 + v_m)/2$.}
\label{triangle}
\end{figure*}

\begin{figure*}[ht!]
\begin{center}
\begin{tabular}{ccc}
\includegraphics[angle=0,width=0.30\textwidth]{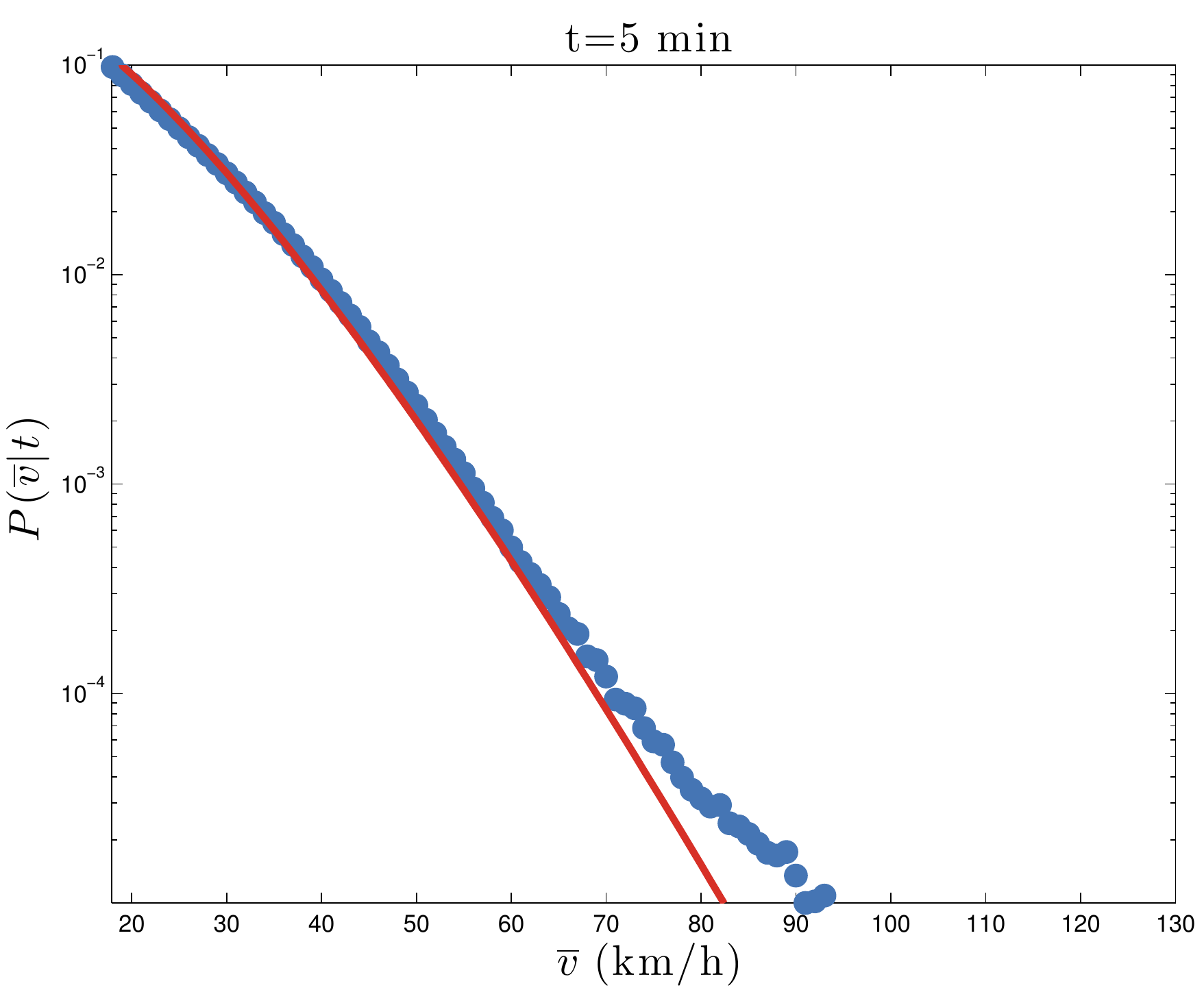}&
\includegraphics[angle=0,width=0.30\textwidth]{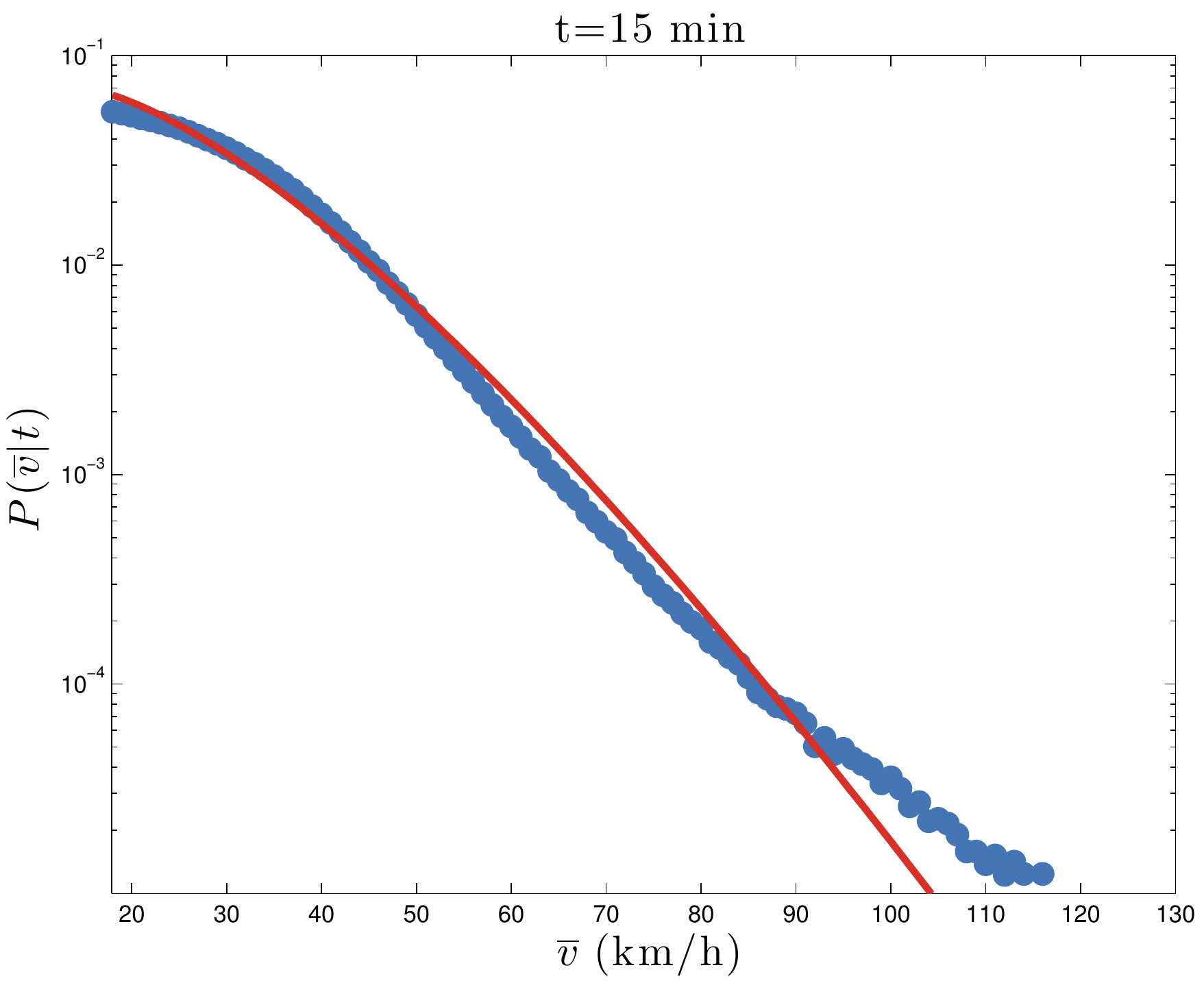}&
\includegraphics[angle=0,width=0.30\textwidth]{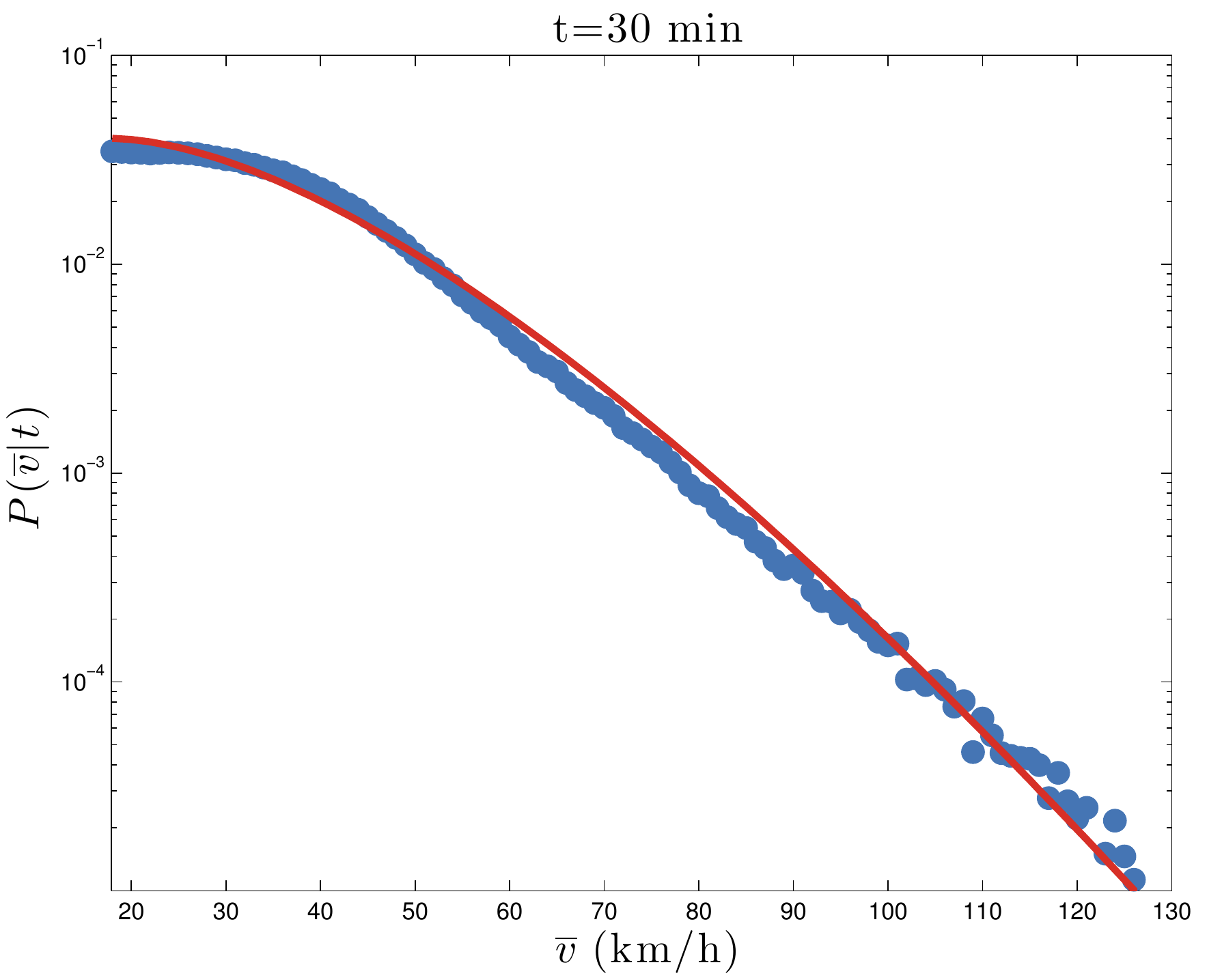}\\
\includegraphics[angle=0,width=0.30\textwidth]{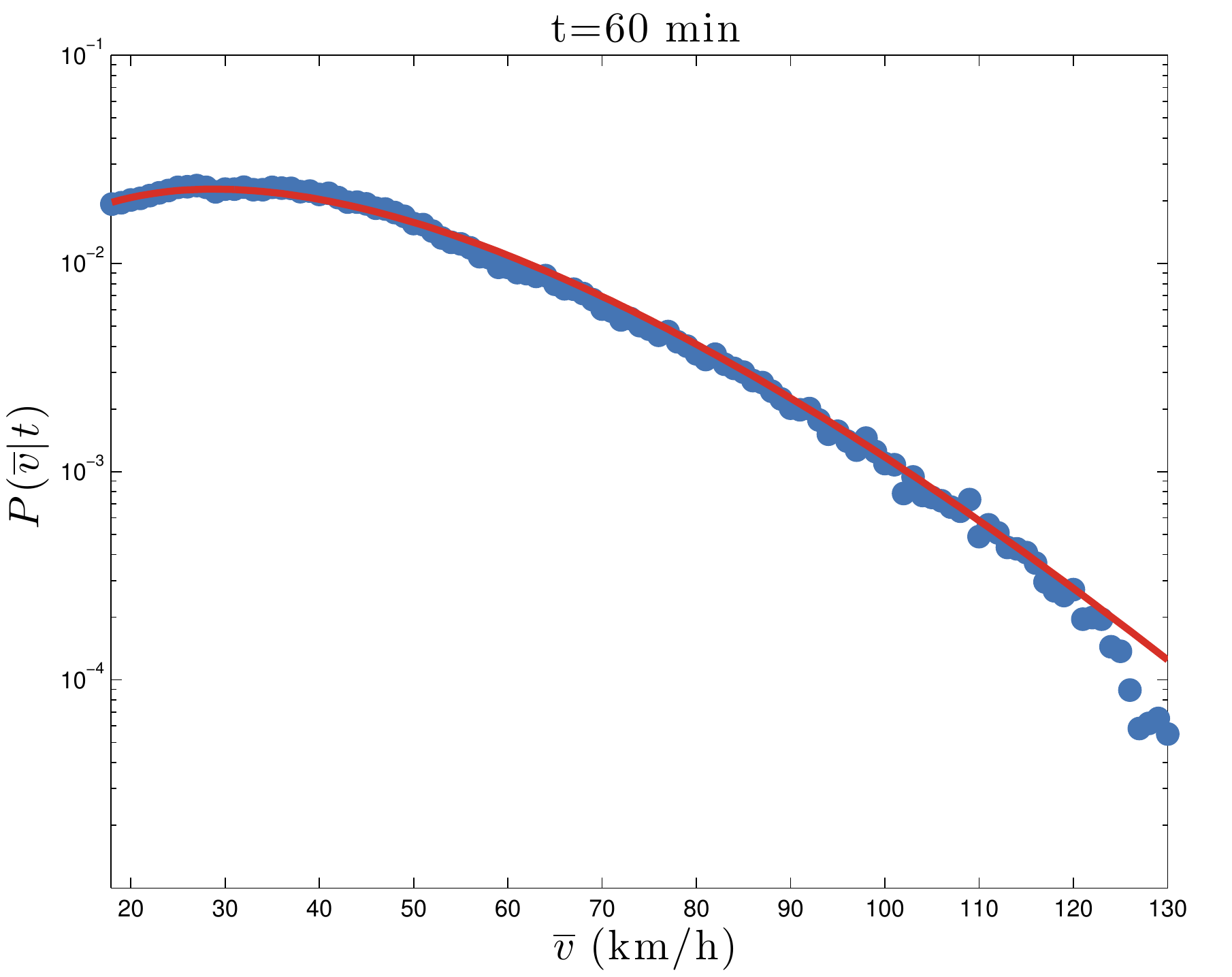}&
\includegraphics[angle=0,width=0.30\textwidth]{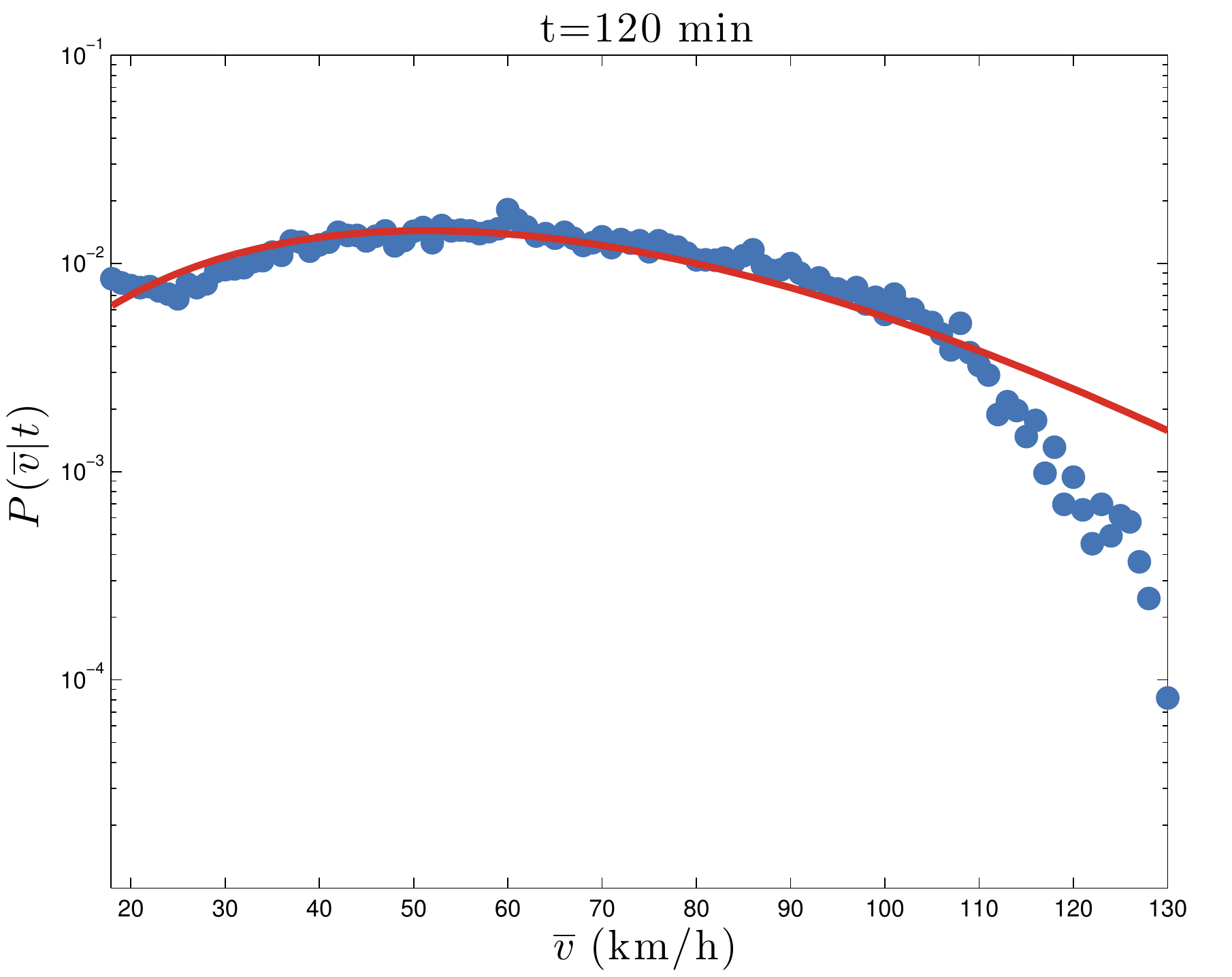}&
\includegraphics[angle=0,width=0.30\textwidth]{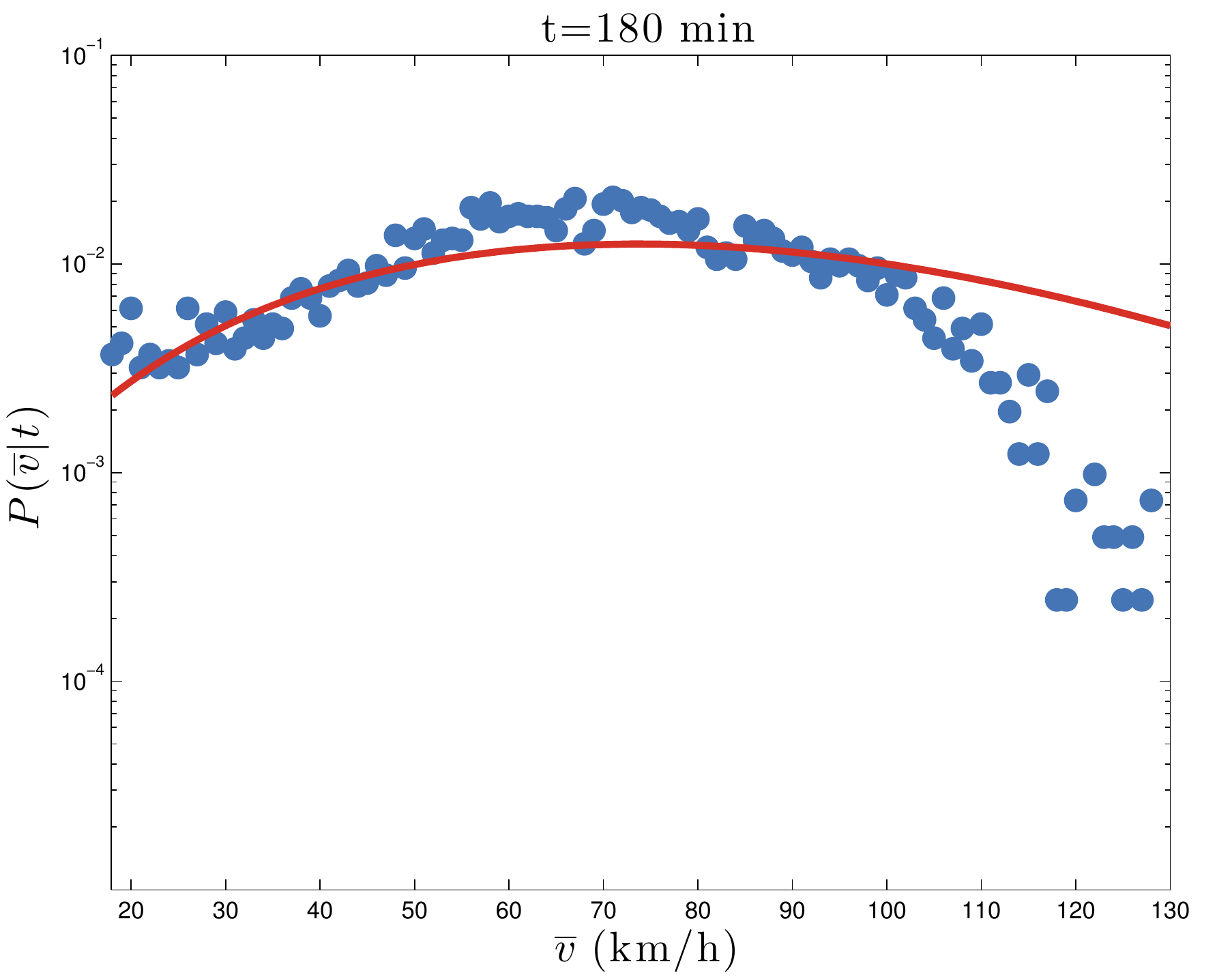}\\
\end{tabular}
\end{center}
\caption{{\bf Conditional distribution $P(\overline{v}|t)$ of average velocities at different travel times.} We fit the curve $P(\overline v|t)$  in the interval $ [v_0,v_\mathrm{max}]$, with $v_0 = 17.9$ km/h and $v_\mathrm{max} = 130$ km/h. We fit simultaneously all the curves for $t = 5, 10, 15, 20, \dots , 180$ minutes, and we show here six examples represented by blue dots. The best fit curve, represented in red, is given by Eq.~(\ref{Pv}) with $p' = 1.06$ jumps/h and $\delta v' = 20.9$ km/h.}
\label{speedDist}
\end{figure*}

In this simple model, we assume that the transportation network has $n$ layers $L_n$, corresponding to different travel speeds $v_n$ (see the schematic representation in Fig.~\ref{triangle}). For the sake of simplicity, we assume that the velocity differences between layers are constant and equal to  $\delta v$. An individual starts his trip of duration $t$ in $L_0$, with base speed $v_0$, while on layer $L_k$, he is traveling faster at speed $v_k = v_0 + k \delta v$. We assume that a trip is composed of two phases. In the first one, the trajectory progressively jumps from slower to faster transportation modes. In the second phase, there is the inverse process where the trajectory progressively jumps down the layers until it finally reaches the base layer at time $t$. We also assume that the process has a Poissonian character, where all individuals have the same probability per unit time $p$ to jump to the successive layer and to change their velocity (both in the ascending and descending phases). Together with the last assumption that on average the duration of both phases are equal and neglecting any saturation issue (see Methods), we can: i) estimate the maximal speed $v_m = v_0 + k(t_m) \delta v$, where $k(t_m)$ is the number of jumps at mid trajectory $t_m = t/2$; ii) approximate the average speed $\overline{v}=1/t\int_0^{t} v(t)\mathrm{d}v = (v_0+v_m)/2$ (see Fig.~\ref{triangle}). Since the process is Poissonian, $\langle k(t_m) \rangle = pt_m$ and the average speed is then given by
\begin{equation}
\langle \overline v(t) \rangle  = v_0 + \frac{p \delta v}{4} t
\label{eq:acceleration}
\end{equation}
where the brackets denote the average over the Poisson variable $k$. The average speed thus grows linearly with $t$ before reaching the saturation imposed by the finite number of layers. Remarkably enough, this simple model allows us also to predict the shape of the conditional probability distribution $P(\overline v|t)$. Indeed, the number of jumps $k$ is distributed following the Poisson distribution $P(k) = \frac{e^{-\lambda} \lambda^k}{k!}$ with $\lambda = pt/2$. Using the Gamma function as the natural analytic continuation of the factorial $k!=\Gamma(1+k)$, we obtain the distribution
\begin{equation}
P(\overline v|t)  =  \frac{1}{\delta v'} \frac{\exp\left(-p't + \frac{\overline v-v_0}{\delta v'}\log(p't) \right)}{\Gamma(1+\frac{\overline v-v_0}{\delta v'})}
\label{Pv}
\end{equation}
where $p' = p/2$ and $\delta v' = \delta v/2$. In Fig~\ref{speedDist} we show that the empirical distribution of the velocities at fixed time $P(\overline v|t)$ for travel-times ranging between $5$ and $180$ minutes are consistent with Eq.~(\ref{Pv}) with $\overline t = 0.30$h, $\delta v' = 20.9$ Km/h, $p' = 1.06$ jumps/h and $v_0 = 17.9$ km/h. This makes the velocity gaps $\delta v\approx 40$ km/h, consistently with the progression of the most common speed limits in Italy: 50 km/h (urban), 90 km/h  (extra-urban), 130 km/h (highways). This result suggests that a multilayer hierarchical transportation infrastructure can explain the constant acceleration observed in both public and private transportation. This model also allows to estimate the base speed $v_0$ as the intercept value in Fig.~\ref{speeds} (b) and (d), and the acceleration $a$ is expected to be proportional to the probability of jump to faster layers $p$ and to the gap between layers $\delta v$ (see Methods).

\subsection*{Displacement distribution}

\begin{figure*}[ht!]
\begin{center}
\begin{tabular}{cc}
\includegraphics[angle=0,width=0.45\textwidth]{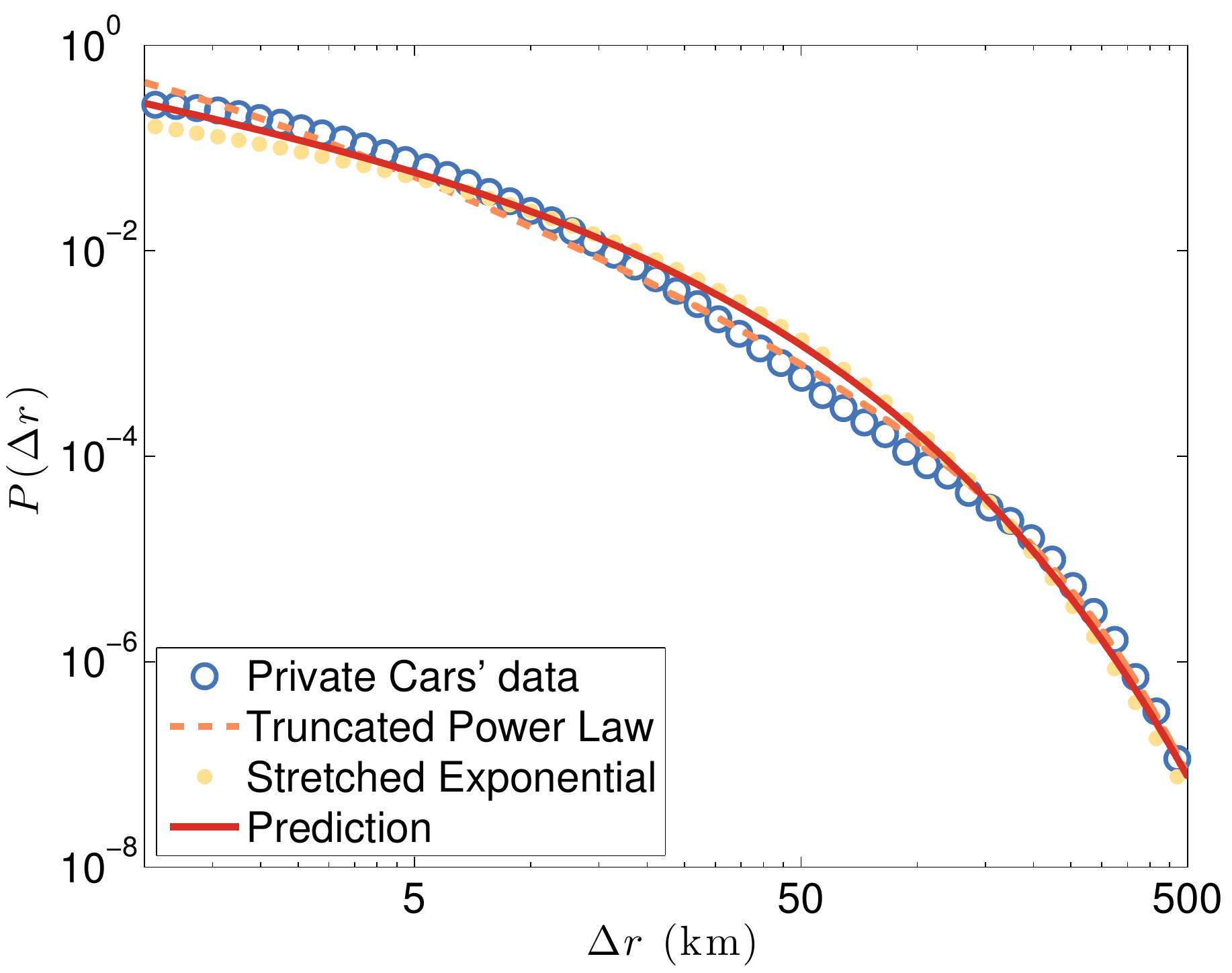}&
\includegraphics[angle=0,width=0.45\textwidth]{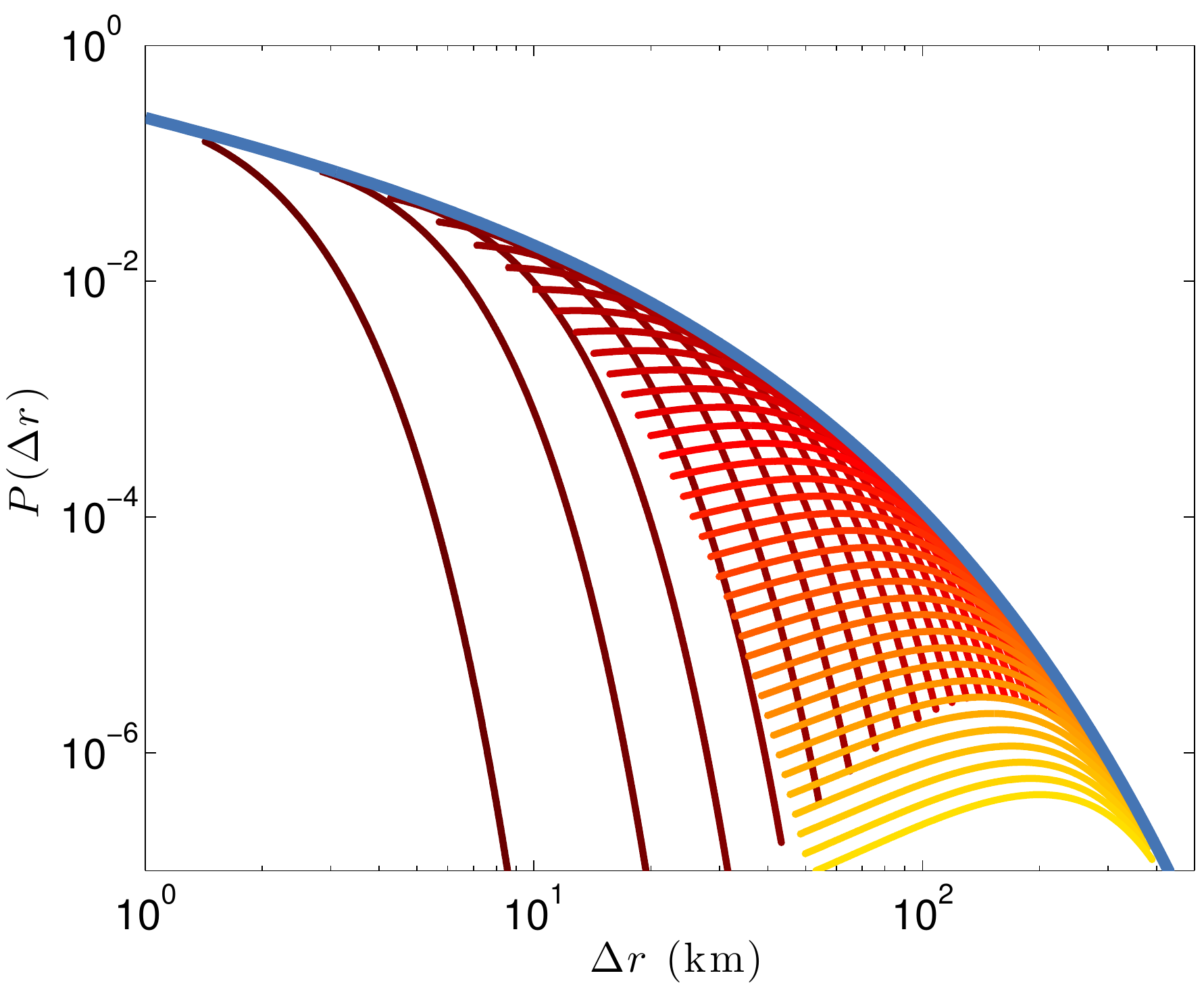}\\
\end{tabular}
\end{center}
\caption{{\bf Distribution of displacements (Left) }. We show (blue circle) the aggregated empirical distribution $P(\Delta r)$ for all the $\approx 780,000$ cars in our dataset. The orange dashed line represents a fit with a truncated power law of the form $P(\Delta r)\propto (\Delta r + \Delta r_0)^{-\beta} \exp(-\Delta r/\kappa)$ with $\beta = 1.67$, $\kappa= 84.1$ km, $\Delta r_0 = 0.63$ km. The coefficients of the fit are close to those found for mobile phone data in the US~\cite{Gonzalez:2008}.
The yellow dots represents the fit with the stretched exponential Eq.~(\ref{randomAccPr}) ($C= 0.63 \mathrm{km}^{-0.5}$) associated to uncorrelated accelerations model. The red solid line shows the prediction based on our model Eq.~(\ref{eq:dr}) with  i) $v_0=17.9$ km/h estimated as the intercept of the fit in Fig.~\ref{speeds}; ii)  $p'= 1.06$ jumps/h, and $\delta v = 20.9$ km/h estimated from the fit of $P(\overline v|t)$ in figure~\ref{speedDist}; iii) $\langle t\rangle = 0.30$ h coming from the  average of the travel-times for all selected trips (see Methods). This prediction is of remarkable quality that can reasonably compared with the commonly used direct fit with a truncated power law with 3 free parameters.
{\bf (Right)} Visual representation of the integral in Eq.~(\ref{eq:dr}). The $P(\Delta r)$ (blue line) is obtained as the superimposition of Poisson distributions of Eq.~(\ref{Pv}) with different $\lambda = p't$. We display the distribution for $t = 5,10,15,\dots,180$ minutes with a colormap going from red to yellow. The domain of these curves is limited by velocities between $v_0$ and 130 km/h. The area under each curve is decreasing with time as $\propto \mathrm e^{-t/\overline t}$.
}
\label{distances}
\end{figure*}

Finally, the shape of the displacement distribution $P(\Delta r)$ can be computed as a superimposition of Poisson distributions (see Fig.~\ref{distances} right)
\begin{align}
\nonumber
P(\Delta r) & = \int_0^\infty dt \int_{v_{1}}^{v_{2}} d\overline v \delta(\Delta r - \overline vt)P(t)P(\overline v|t) \\ 
& = \int_0^\infty \frac{dt}{\overline t t} \frac{\exp\left(-\left(p'+\frac{1}{\overline t}\right) t + \log(p't) \frac{\Delta r/t-v_0}{\delta v'} \right)}{\Delta v'\Gamma\left(1+\frac{\Delta r/t-v_0}{\delta v'}\right)} 
\label{eq:dr}
\end{align}
where $\delta$ is the Dirac delta function. The exact form for Eq.~(\ref{eq:dr}) cannot be analytically computed but the limiting behavior of for large $\Delta r$ is again Eq.~(\ref{randomAccPr}) (see Supplementary information). In particular, this distribution is not broad, in clear contrast with L\'evy flights which have divergent moments and are governed by large fluctuations. Since the distribution $P(\Delta r)$ is not a fat tail distribution, all the phenomena associated to L\'evy flights, such as super-diffusion for example, are not expected for randomly accelerated walks. In Fig.~\ref{distances} left, we show that our prediction on the shape of $P(\Delta r)$ is consistent with the empirical data. Similar results can be obtained studying the mobility of single cities (see Fig.~\ref{rCities}). We stress that the curve we propose is not an a posteriori best fit of the empirical $P(\Delta r)$ but the curve predicted from Eq.~(\ref{eq:dr}) knowing the average travel-time $\overline t$ and using the values $v_0$, $p'$, $\delta v'$ used in the description of the empirical $P(\overline v|t)$.

The fact that we can fit the data by different forms, as the (truncated) power law fit proposed in various studies~\cite{Brockmann:2006,Gonzalez:2008} (see Table~\ref{table1})  or the stretched exponential of Eq.~(\ref{randomAccPr}), is an illustration of the difficulty to extract mechanistic information from empirical data only using macroscopic statistical laws, without taking into account the dynamical properties of the underlying processes. In particular, urban mobility seems thus not to be a L\'evy process as inferred from truncated power law fits, but its random nature is governed by transitions between modes or roads with different velocities. In contrast with purely empirical works that led to L\'evy flight models, our model is based on the interaction between individuals and the transportation networks~\cite{Benhamou:2007}. Our approach accounts for the displacement distributions, but also provide a possible explanation of the processes generating such macroscopic patterns.



\clearpage

\section*{Methods}
\subsection*{Private Transportation Data}

We compute spatial displacements and travel times for private
transportation from a database of GPS measures describing the
trajectories of private vehicles in the whole Italy during the month
of May 2011. This database is mainly set up by private vehicles, since Taxi
or delivery companies use their own GPS systems and do not
contribute to the database. A small percentage of vehicles belongs to
private companies and are used for professional reasons. This database
includes $\approx 2\%$ of the vehicles registered in Italy, containing
a total of $128,363,000$ trips performed by $779,000$ vehicles. Records
contain information about engine starts and stops, and travel-times also include the eventual time spent looking for parking. When the quality of the satellite signal is good, we have an
average spatial accuracy of the order of $10$ meters, but in some cases it can
reach values up to $30$ meters or more~\cite{Bazzani:2011}. The temporal
resolution is of the order of the second.


We have applied correction and filtering procedures to exclude from
our analysis the data affected by systematic errors. Approximately $10\%$ of
the data were discarded for this reason. When the engine is switched on or
when the vehicle is parked inside a building, there are some errors
due to the signal loss. In such cases, we use the redundant information
given by the previous stopping point to correct $20\%$ of the data
When the engine's was off for less than 30 seconds, the subsequent
trajectory is considered as a continuation of the same trip if it is
not going back towards the origin of the first trajectory.


For privacy reasons, the drivers' city of residence is
unknown. Therefore, it has been necessary to associate each car to an
urban area using the available information. We do that identifying a
driver as living in a certain city if the most part of its parking
time was spent in the corresponding municipality area. Then, for each
driver we have considered all the mobility performed, both within and
outside the urban area.

\subsection*{Details of the model}

The model proposed here assumes that urban mobility is performed at an (euclidean) speed of  $v_0 \approx 18$ km/h. Very short trips hardly reach the base speed $v_0$ we impose in our model. For this reason, in all the measures proposed in this paper, we considered only trips longer than $1$ km and $5$ minutes. 

The upper bound to 130 km/h limits the number of jumps to $k = (130 \mathrm{km/h} - v_0)/\delta v \approx 3$. We a jumping rate of order $p'\approx 1$ jump/h, we may expect significant deviations from our prediction for trips longer than $4 hours$. For long trips one cannot in principle approximate the area below the step function in Fig.~\ref{triangle} with a triangle. Indeed, the Poisson fit in Fig.\ref{speedDist} overestimates the frequency of very fast trips for $t= 120$ and $180$ minutes.  If the acceleration kicks are limited by a finite number of layers, the speed grows linear for small times and converge asymptotically to a limiting speed (see Supplementary Information).


\section*{Acknowledgments}
We thank P. Krapivsky, M. Lenormand and R. Louf for useful discussions. 
We thank Octo Telematics S.p.A. for providing the GPS database.

\section*{Notes}
RG and MB designed research, performed research and wrote the
paper. AB and SR obtained the dataset and performed the data
pre-elaboration. RG prepared the figures.


\setcounter{figure}{0}
\renewcommand{\thefigure}{S\arabic{figure}}

\setcounter{table}{0}
\renewcommand{\thetable}{S\arabic{table}}

\clearpage
\section*{Supplementary information}

\

\subsection*{Trip duration in public and private transportation}

\begin{figure*}[ht!]
\begin{center}
\begin{tabular}{cc}
\includegraphics[angle=0,width=0.45\textwidth]{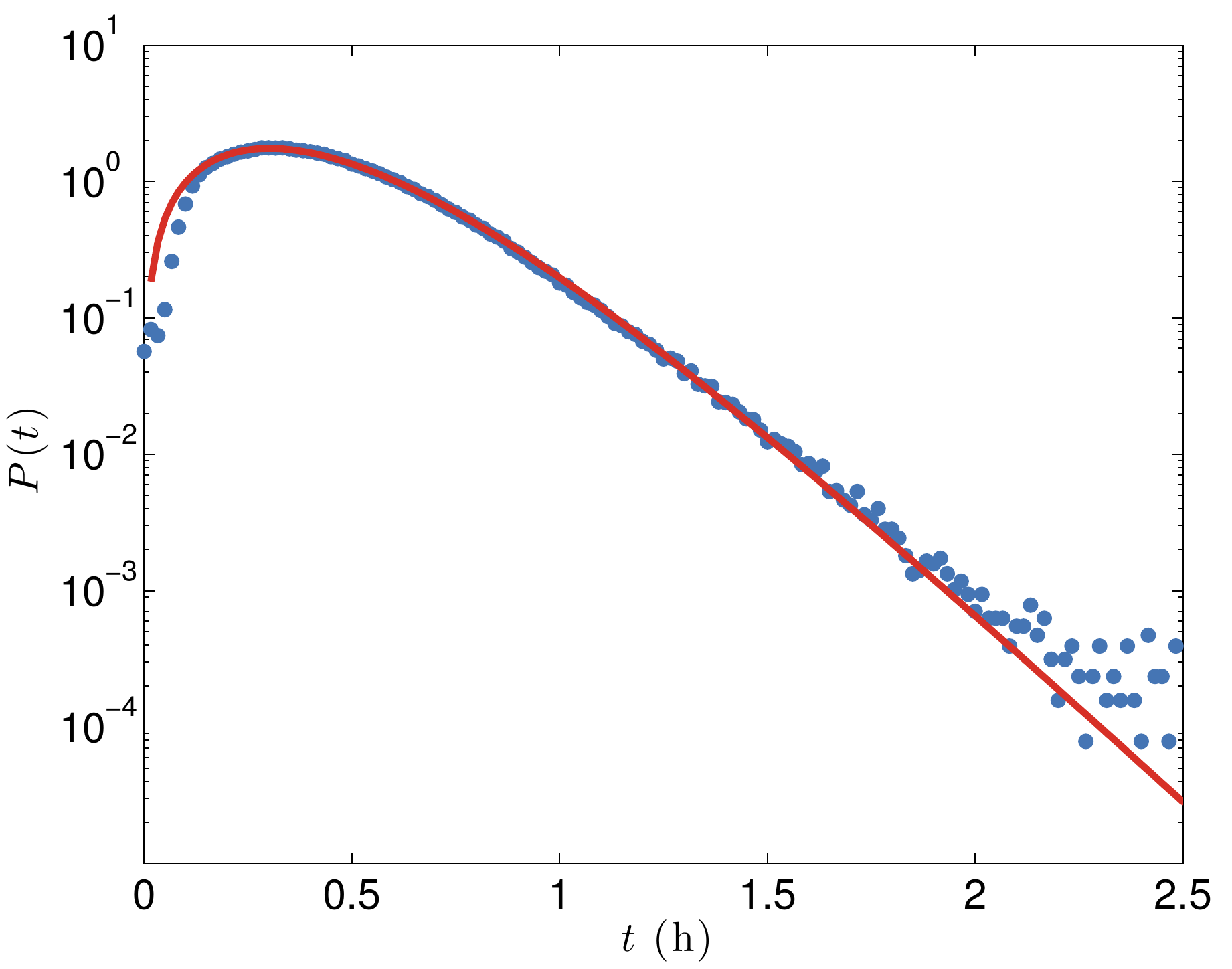}&
\includegraphics[angle=0,width=0.45\textwidth]{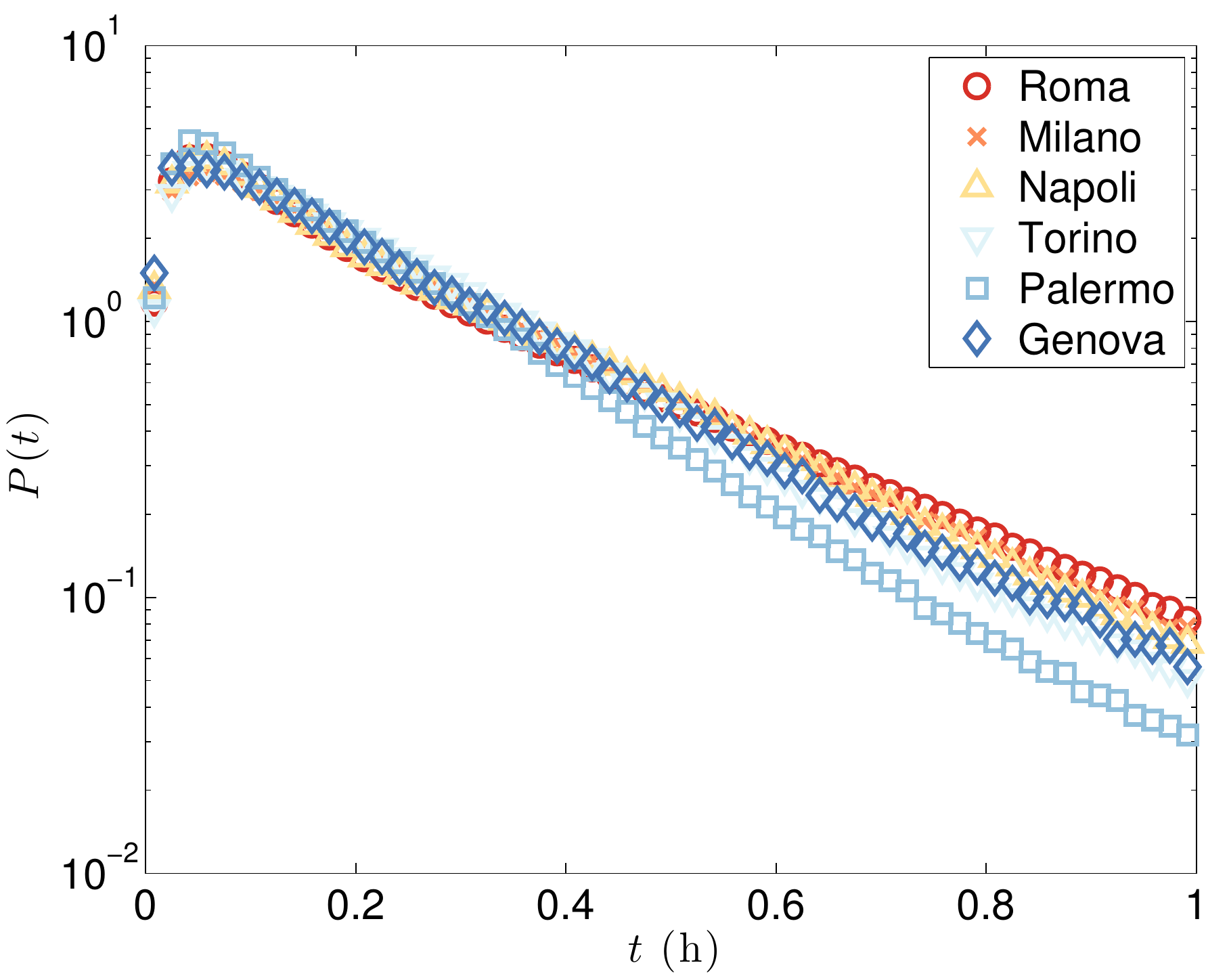}\\
\end{tabular}
\end{center}
\caption{{\bf Travel-times distributions in different cities and modes
    of transport.} {\bf(Left)} Travel-times extracted from a 5\%
  sample of all Oyster card journeys performed in a week during
  November 2009 on bus, Tube, DLR and London Overground. Data available
  at https://api-portal.tfl.gov.uk/docs (accessed 3/9/2015). The
  probability distribution is fitted with the curve
  $P(t) \propto (1-\exp(-t/c)) \exp(-c\exp(-t/c)/b - t/b)$ proposed
  in~\cite{Gallotti:2015ttb} for the description of daily travel time
  expenditures. The parameters $c= 36\pm2$ minutes and $b=9.3\pm0.3$ minutes
  show that the average travel-time given by the exponential tail
  is of order of $9$ minutes, but at least $36$ minutes are needed for
  reaching this exponential behavior. {\bf(Right)} In contrast with Fig.~1,
  where the travel-times of private cars were normalized, here we
  display the differences between the distributions for the 6 largest
  cities in Italy. The tail appears to be exponential for travel-times
  shorter than the classical value of daily travel-time budget of 1
  hour~\cite{Metz:2008,Gallotti:2015ttb}.  }
\label{figS1}
\end{figure*}

\begin{figure*}[ht!]
\begin{center}
\includegraphics[angle=0,width=0.45\textwidth]{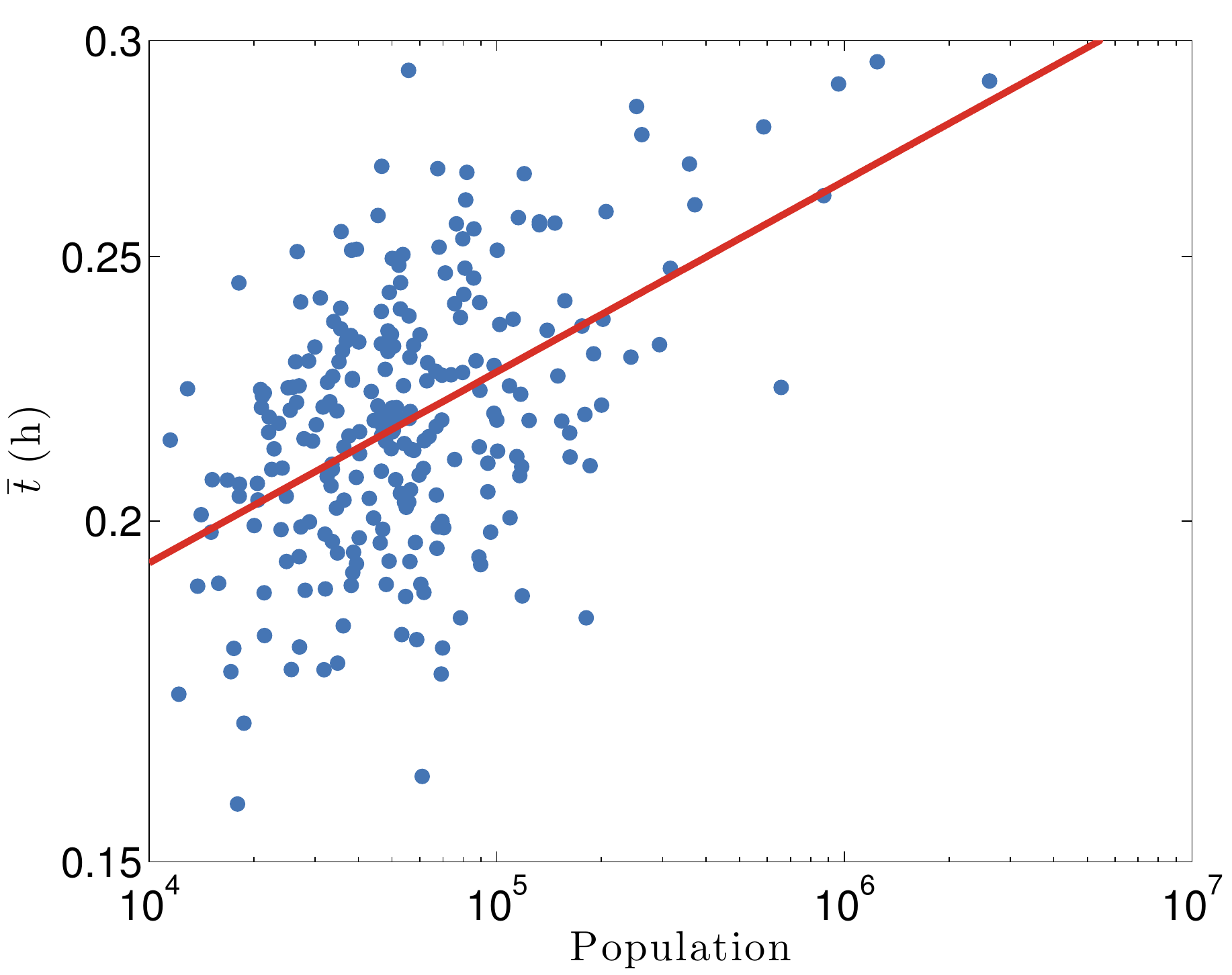}
\end{center}
\caption{{\bf Variation of the average travel-time of private cars
    across Italian cities.} We display here the values for all cities
  with at last $500$ drivers that are monitored. The average
  travel-time (computed here over all trips in a given city) is in the approximate range $[9,18]$ minutes. These
  values are correlated $r = 0.40$ with the city population. The solid
  line represents a fit for a power law $\overline t  \propto$
  population$^\gamma$ with $\gamma= 0.07\pm0.02$.}
\label{figS2}
\end{figure*}

\subsection*{Dependence of velocities over time in public transportation.}

\begin{figure*} [ht!]
\begin{center}
\begin{tabular}{cc}
\includegraphics[angle=0, width=0.45\textwidth]{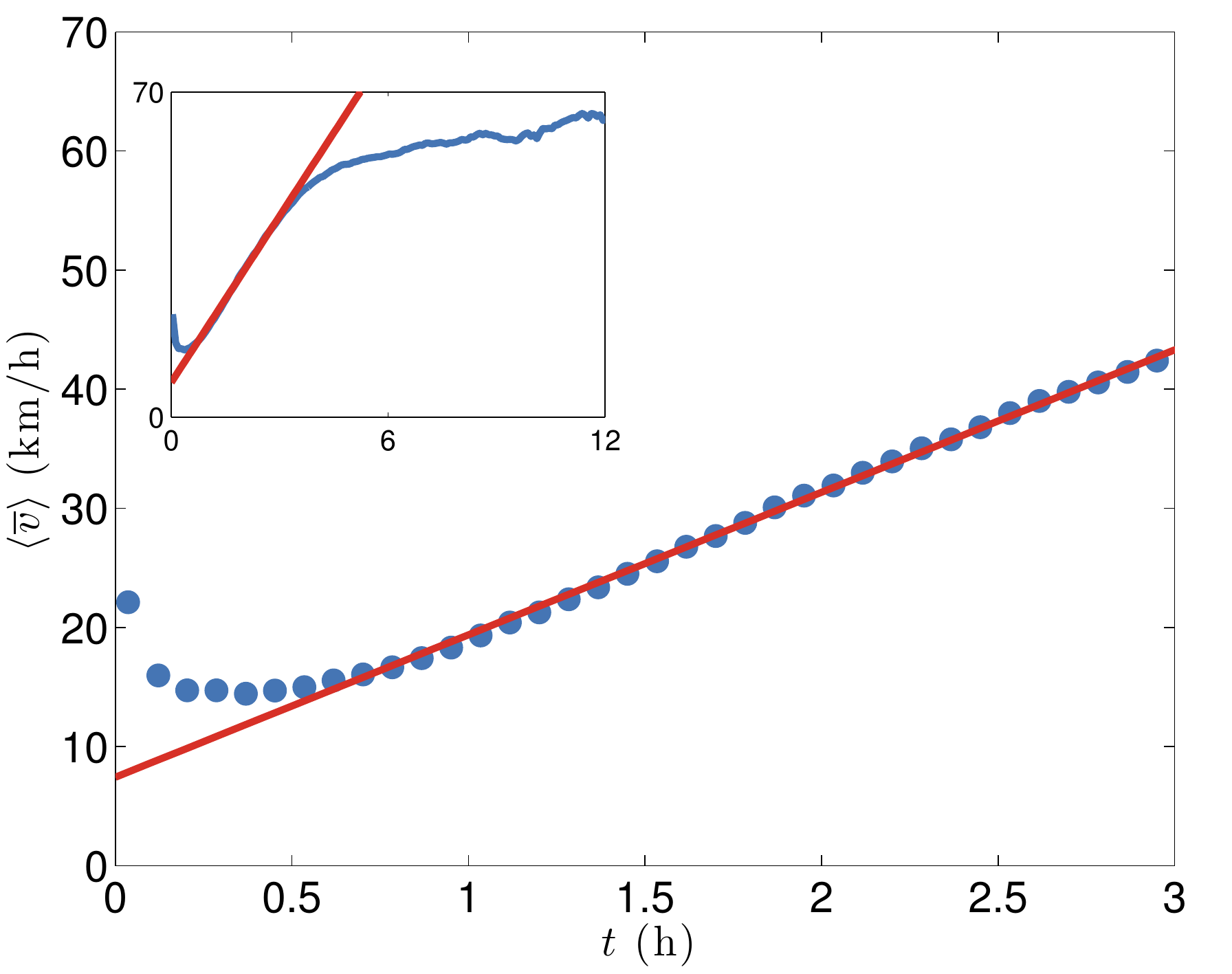} &
\includegraphics[angle=0, width=0.45\textwidth]{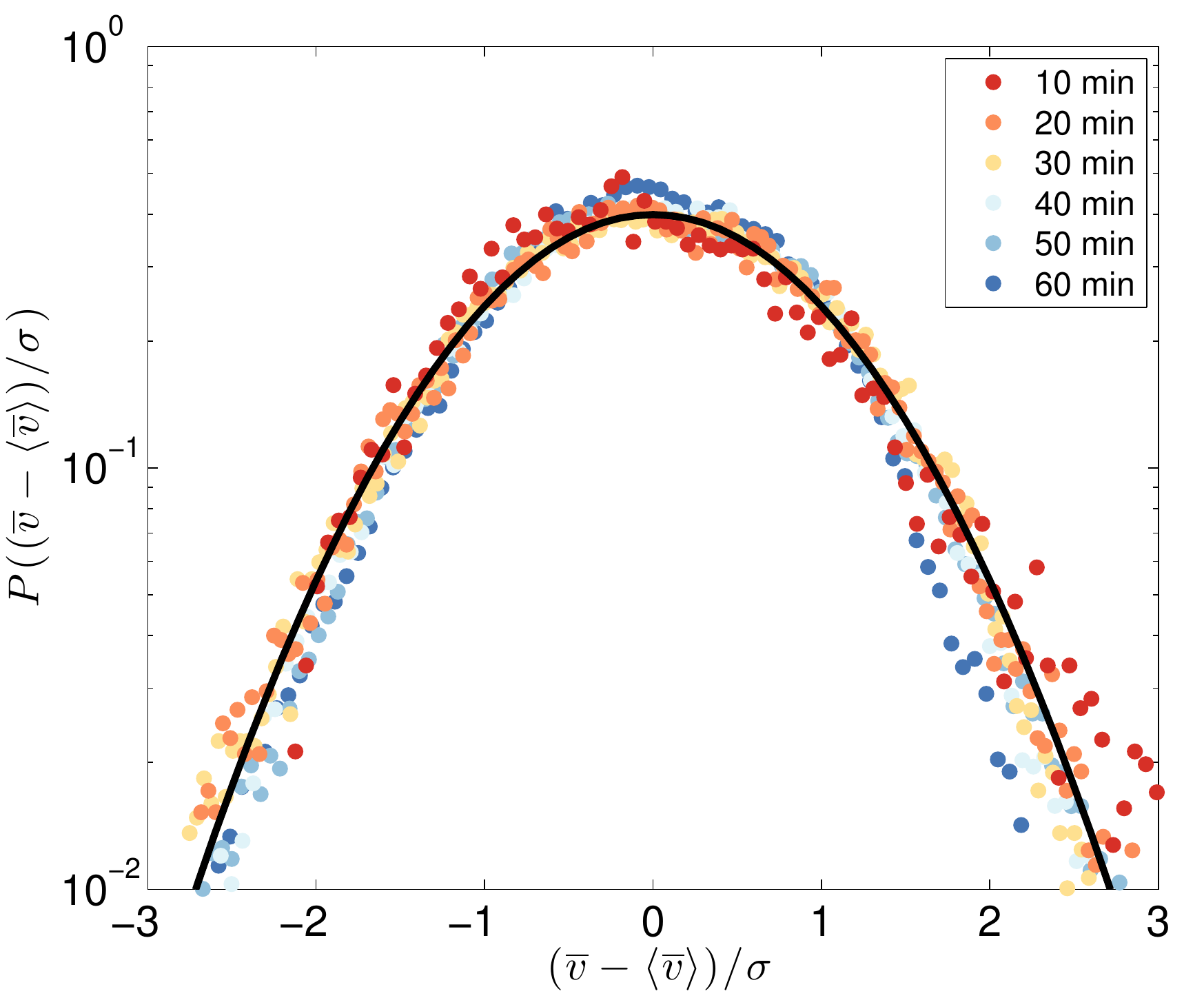} \\
\end{tabular}
\end{center}
\caption{
{\bf Acceleration of public transportation. (Left)} We observe also for public transportation trajectories (linking all possible origins and destinations in Great Britain) a uniform acceleration for trips of duration between 30 minutes and 3.5 hours. The solid red line represents the fit with $v_0 = 7.4\pm0.4$ km/h and $a = 12.0\pm0.2$ km/h$^2$. In this case, very short trips appear as faster thanks to the likely absence of connections.  {\bf (Right)} Conditioned speed distribution for trajectories originated in Charing Cross in London. Selecting trips of duration shorter than 1 hour we have manifestly a normal distribution. 
}
\label{publicTransport}
\end{figure*}

We estimate velocities for public transportation trips from an open dataset providing a complete snapshot of the multilayer temporal network of public transport in Great Britain in October 2010~\cite{Gallotti:2015mtn}.  This dataset assumes a waiting time before a flight of 2 hours, a waiting time after a flight of 30 minutes and, where not explicitly defined by the transportation agencies, a walking distance of $250$ m. The starting time in our analysis of a week is Monday, 8am.  Starting from that time, we consider for all trips from a node $i$ to a node $j$, the time spent in transportation after departing from the node $i$ in any direction. The shortest time-respecting path~\cite{Gallotti:2014} is then identified with a Dijskstra algorithm, where also the time spent walking (at 5 km/h) between two adjacent stops of any modes of transport and the time spent waiting at the connection is integrated to the total travel-time. The euclidean distance between origin and destination is then used for estimating velocities.

Both the averages and the distribution of Fig.~\ref{publicTransport} are not computed on real flows, which are not available with the same spatial extension and definition of the dataset. In this study, we implicitly assumed a uniform travel demand by including all possible origin-destination pairs.

We observe a similar linear growth for the average velocities of trajectories in the public transport system (Fig~\ref{publicTransport} left). In this case we do not have actual individual trajectories but if we assume that there is a uniform travel demand and that travellers make the shortest time-respecting path between origin and destination~\cite{Gallotti:2014}, we can estimate velocities. Short trips, of less than $30$ minutes, tend to be faster thanks to the likely absence of time-consuming connections. For $t\in [0.5,3.5]$ hours, the growth is again linear but both base speed $v_0 = 7.4\pm0.4$ km/h and acceleration $a = 12.0\pm0.2$ km/h$^2$  are smaller than in the case of private transport. In Fig~\ref{publicTransport} right we show that for urban trajectories in London ($t < 1$ h from Charing Cross)  the distribution $P(v)$ is universal and gaussian-like. The difference between the exponential shape of $P(v)$ for private transport and the gaussian-like distribution for public transport might originate in the fact that in public transport the average speed is the weighted average of velocities associated to each edge of the transportation network (plus the effect of waiting time), leading to a normal distribution according to the central limit theorem~\cite{Strano:2015}. In principle, public transports move
according to a deterministic dynamics and arrival times at stations is fixed. However, fluctuations in the average velocity of a trip lead to small delays at different stations and when summed together -for a trip with connections - lead to a Gaussian distribution. 

\subsection*{Accelerated walker with a limited number of layers}

\begin{figure*}[ht!]
\begin{center}
\begin{tabular}{cc}
\includegraphics[angle=0,width=0.45\textwidth]{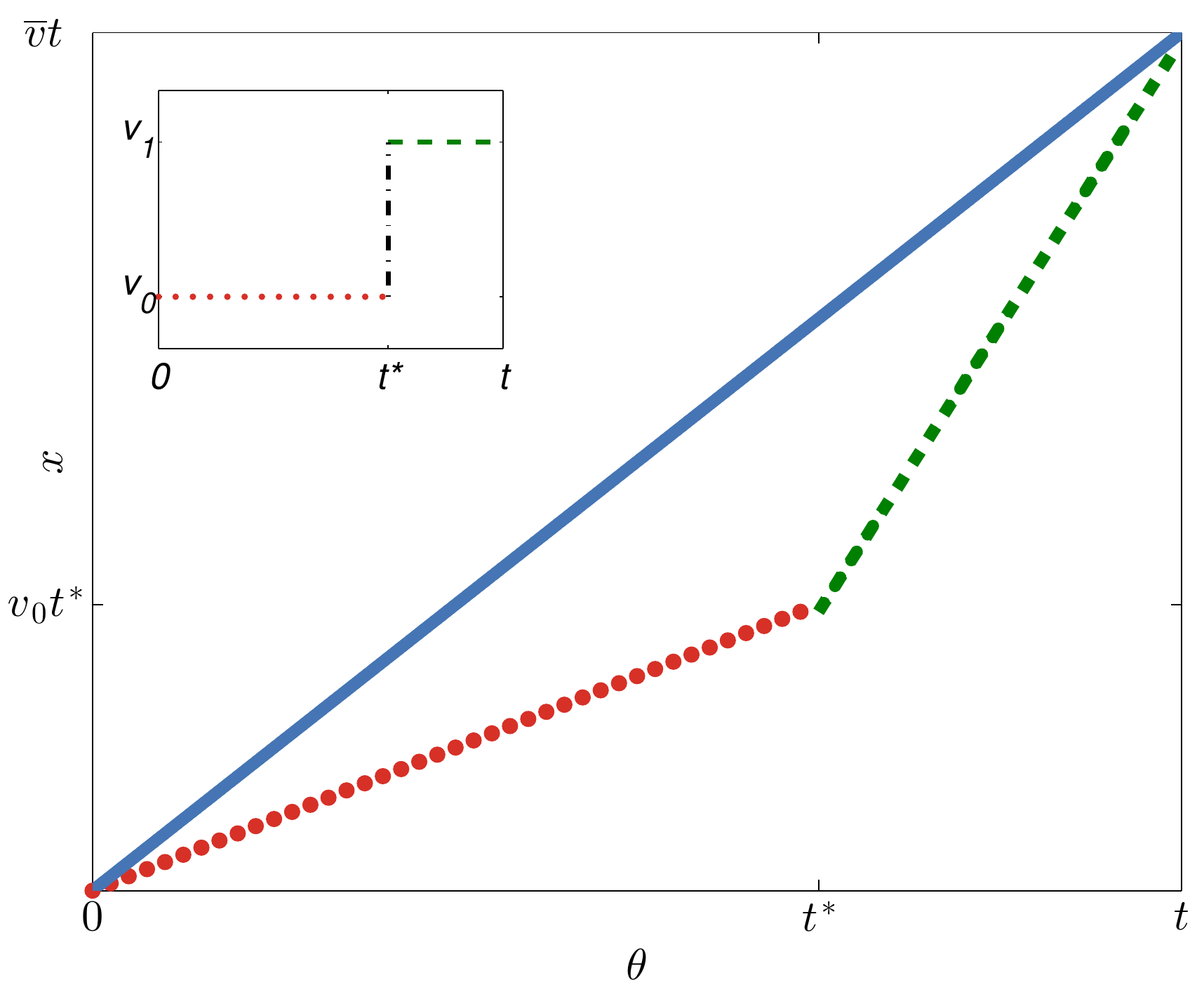}&
\includegraphics[angle=0,width=0.47\textwidth]{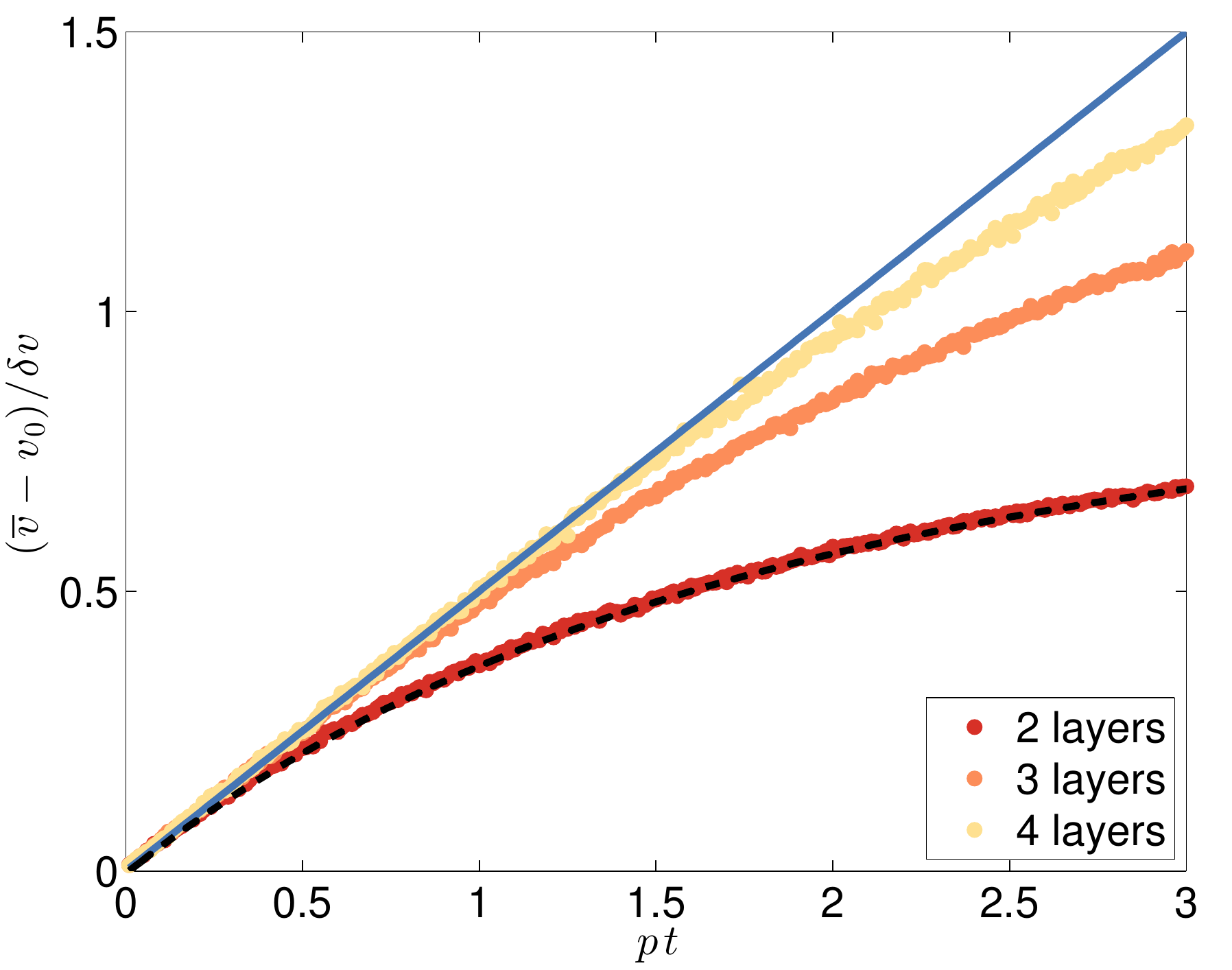}\\
\end{tabular}
\end{center}
\caption{{\bf Average speed in a one-dimensional multilayer
    hierarchical transportation infrastructure.}  {\bf (Left)} The
  dotted line in the plane $(\theta,x)$ represents a trajectory where a
  time $t^*$ is spent on the base layer with speed $v_0$ (red dots)
  and a time $\theta-t^*$ is spent travelling at speed $v_1$ on the fast
  layer (green dashed line). The average speed $\overline v$ is simply
  the weighted average of the two speeds (blue solid line).  {\bf
    (Right)} A numerical simulation confirms the result of
  Eq.~(\ref{eq_vt_teo}) (black dashed line) for the case of two
  layers. We also simulated the case of more than two layers where the
  linear growth regime
  $\overline{v} \approx v_0+\frac{1}{2}p(v_1-v_0)$ (blue solid line)
  extends beyond $pt\ll1$.  }
\label{exactHierarchy}
\end{figure*}

We consider the one dimensional case (the extension to 2d is simple) for a transportation network with only 2 possible layers: $L_0$ and $L_1$ corresponding to different travel velocities. On the layer $L_0$, individuals travel with speed $v_0$, while on $L_1$ they are traveling faster at speed $v_1 = v_0 + dv$. An individual starts its trip of duration $t$ in $L_0$ and has a probability per unit time $p$ to jump to layer $L_1$ and to increase its speed. Being a Poisson process, the probability to jump at time $t^*$ is then given by 
\begin{equation}
P(t^*=\theta)=p\mathrm{e}^{-p\theta}
\end{equation}
and the position at time $t\theta$ of the traveller is 
\begin{equation}
x(\theta)=v_0\theta H(t^*-\theta)+ H(\theta-t^*)\left[v_0t^*+v_1(\theta-t^*)\right]
\end{equation}
where $H(x)$ is the Heaviside function (see Fig.~\ref{exactHierarchy} left).

By averaging the position over $t^*$, we
obtain for the average speed $\overline{v}=\overline{x}/t$ the following simple expression
\begin{equation}
\overline{v}(t)=v_1+(v_0-v_1)\frac{1-e^{-pt}}{pt}
\label{eq_vt_teo}
\end{equation}
(the bar $\overline{\cdot}$ denotes the average over different trips). For the limiting case $pt\ll1$ the average speed grows linearly from the base value $v_0$: $\overline{v}(pt\ll1)\approx v_0+\frac{1}{2}pt(v_1-v_0)$, while for $pt\gg1$ the average speed converges asymptotically to $v_1$: $\overline{v}(pt\gg 1)\approx v_1-(v_1-v_0)/pt$. This simple model thus recovers, in some regime, the linear growth of speed with the duration of the trip, and also the tendency to reach a limiting speed observed in Fig~\ref{speeds}.

If, as in the model we propose in this paper, we have more than two layers with the same jumping probability and speed gap $v_{n+1}-v_{n} = \delta v$ between consecutive layers $L_n$ and $L_{n+1}$, the constant acceleration regime is extended up to $pt>1$ (see Fig.~\ref{exactHierarchy} right). This result suggests that a multilayer hierarchical transportation infrastructure can explain the constant acceleration observed in both public and private transportation. From this model, we can also estimate the base speed $v_0$ with the value of the intercept in the Figures~\ref{speeds},~\ref{publicTransport} left, and \ref{vCities}, while the acceleration $a$ is expected to be proportional to the probability of jump to faster layers $p$ and to the gap between layers $\delta v$.

\subsection*{Variability of $P(\Delta r)$ among cities}


\begin{figure*}[ht!]
\begin{center}
\begin{tabular}{cc}
\includegraphics[angle=0,width=0.45\textwidth]{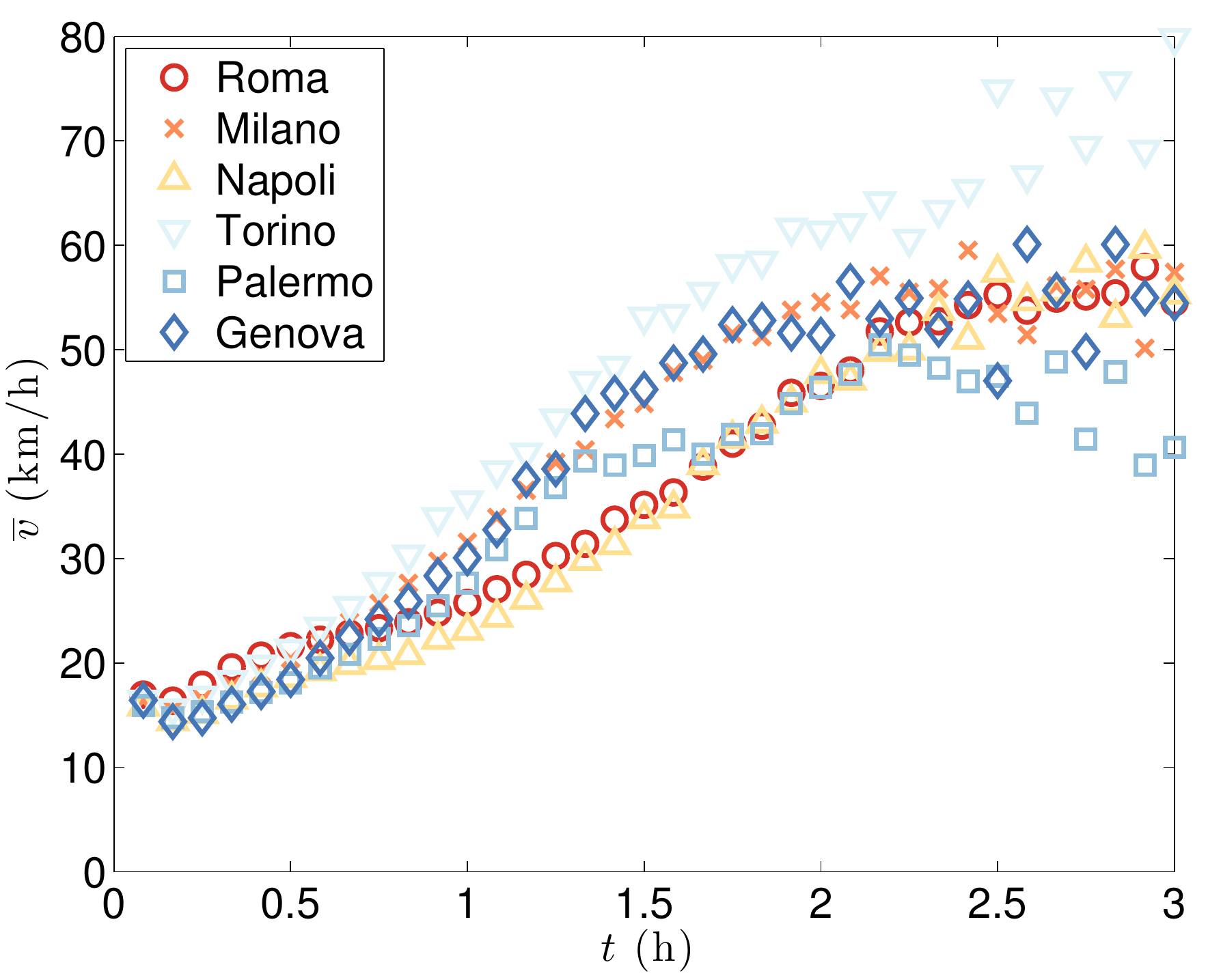}
\end{tabular}
\end{center}
\caption{{\bf Speed growth of cars in the 6 largest Italian cities.} The speed growth for single cities is less uniform than in the aggregate case of Fig.~\ref{speeds}. This is probably related to inhomogeneities of the road network and geographical characteristics of the region around each city.
All curves share a similar value of $\overline v = 16.5\pm 0.5$ km/h for $t = 5$ min, therefore this constant value has been selected as $v_0$ for the prediction proposed in Fig.~\ref{rCities}. }
\label{vCities}
\end{figure*}

\begin{figure*}[ht!]
\begin{center}
\begin{tabular}{cc}
\raisebox{2.3cm}{\rotatebox{90}{(Roma)}}
\includegraphics[angle=0, width=0.4\textwidth]{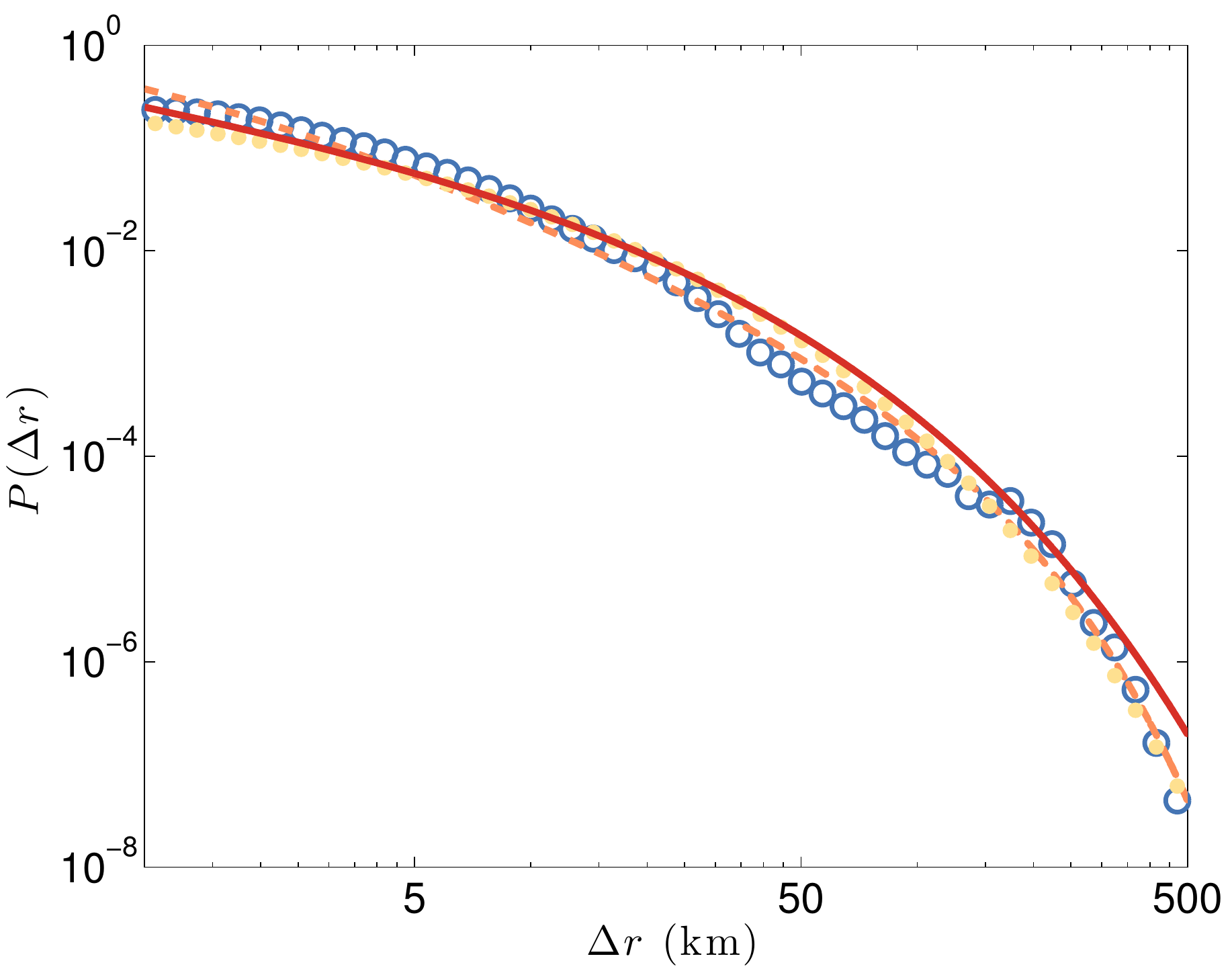}&
\raisebox{2.3cm}{\rotatebox{90}{(Milano)}} \includegraphics[angle=0, width=0.4\textwidth]{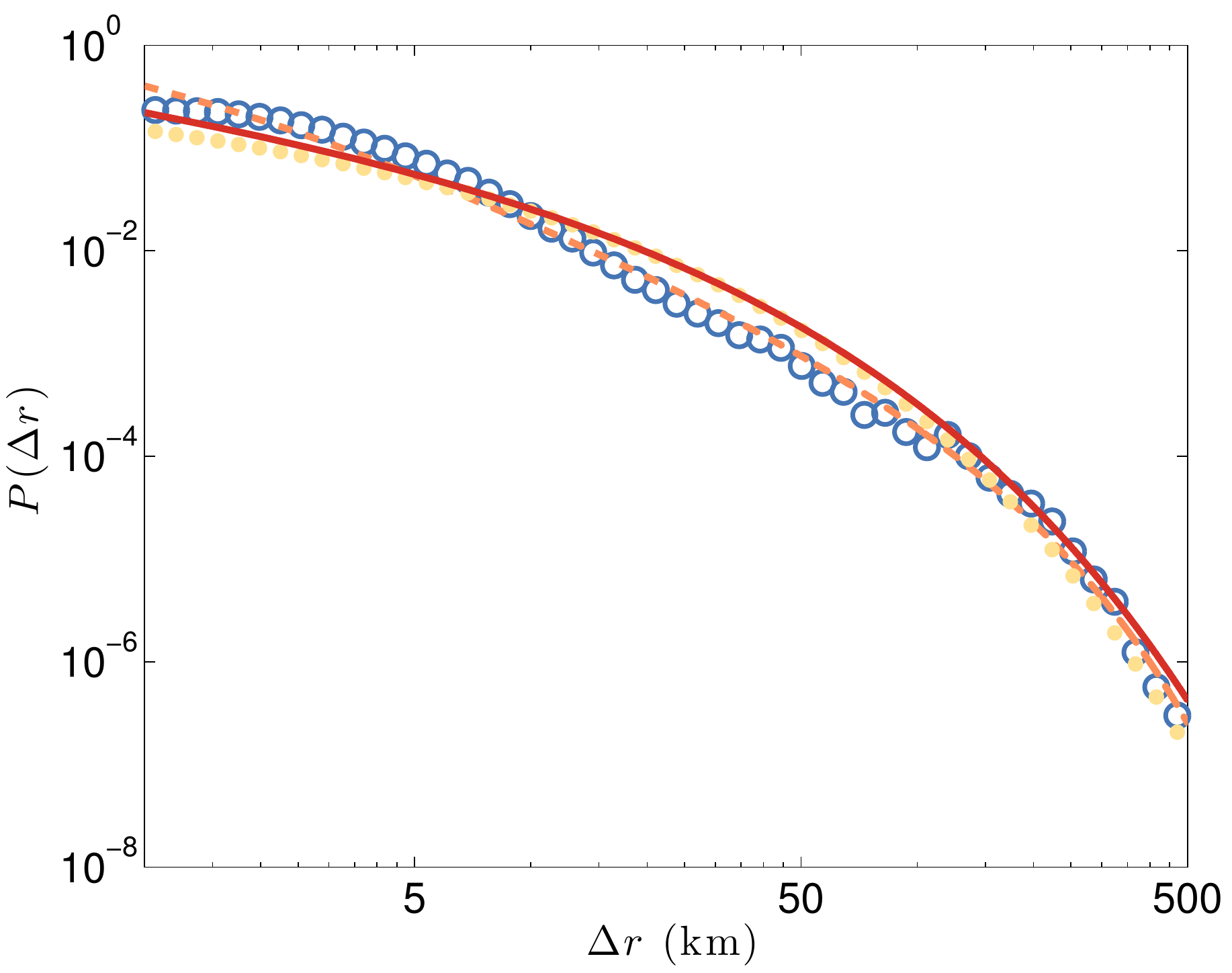}\\
\raisebox{2.3cm}{\rotatebox{90}{(Napoli)}} \includegraphics[angle=0, width=0.4\textwidth]{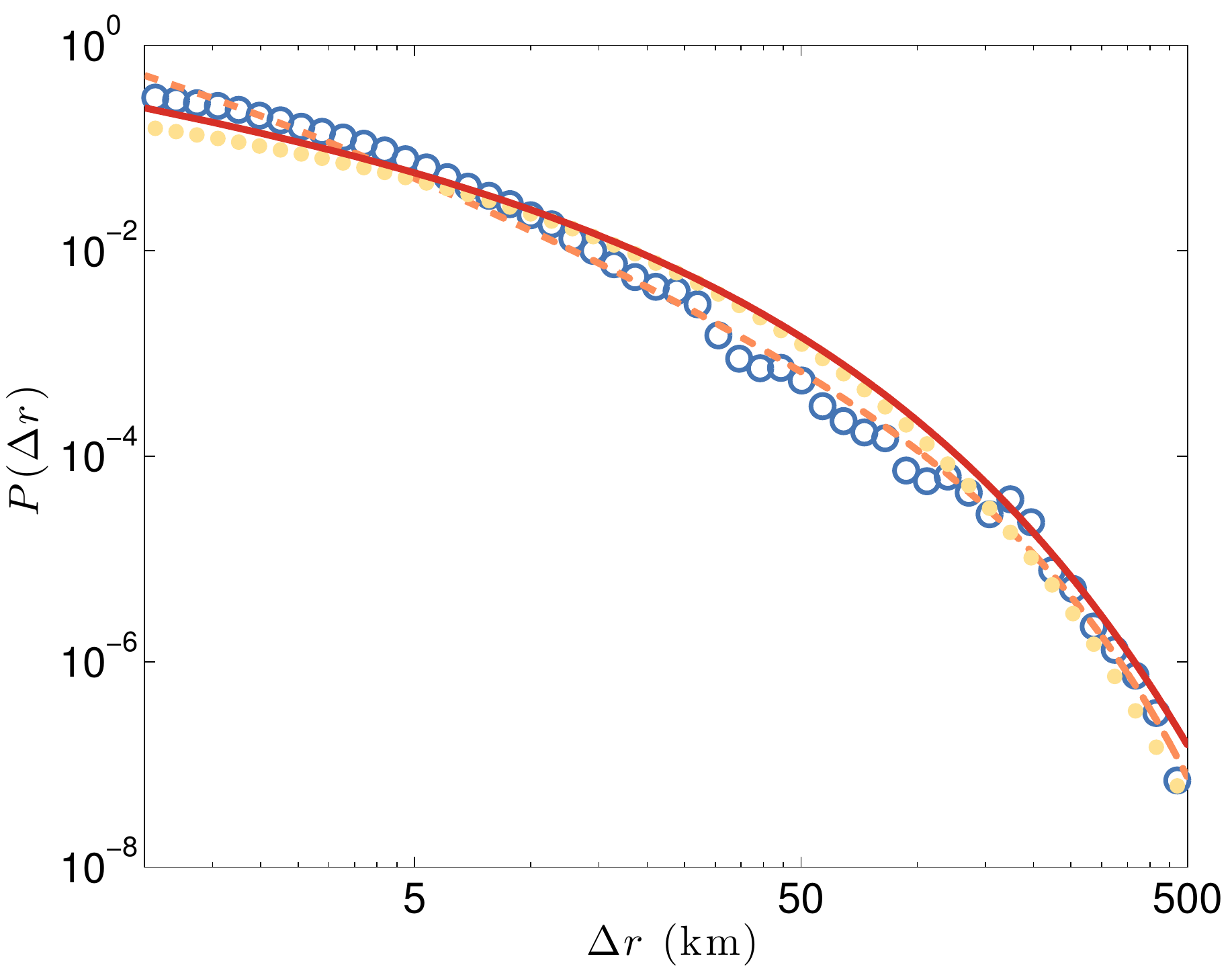} &
\raisebox{2.3cm}{\rotatebox{90}{(Torino)}} \includegraphics[angle=0, width=0.4\textwidth]{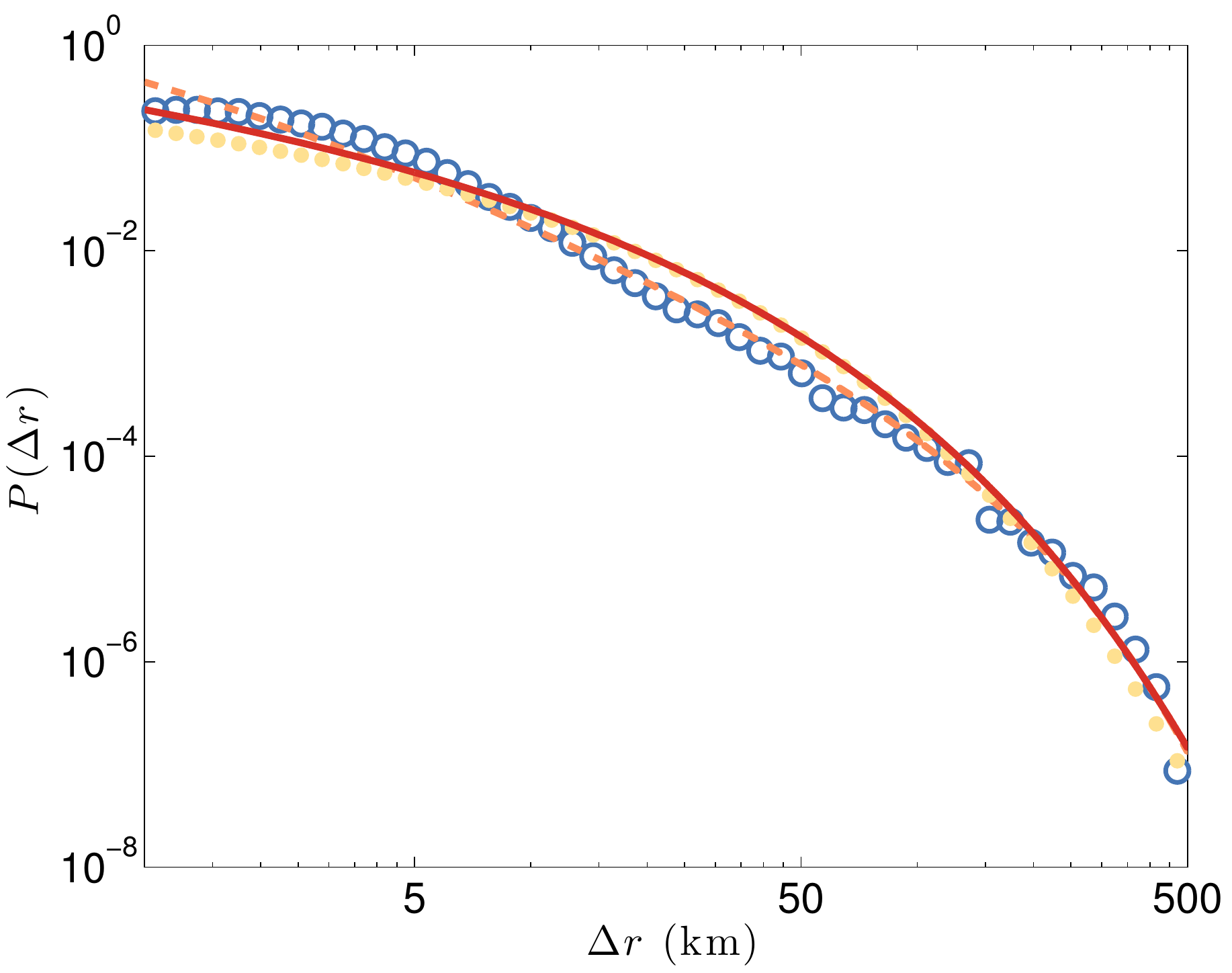} \\
\raisebox{2.3cm}{\rotatebox{90}{(Palermo)}} \includegraphics[angle=0, width=0.4\textwidth]{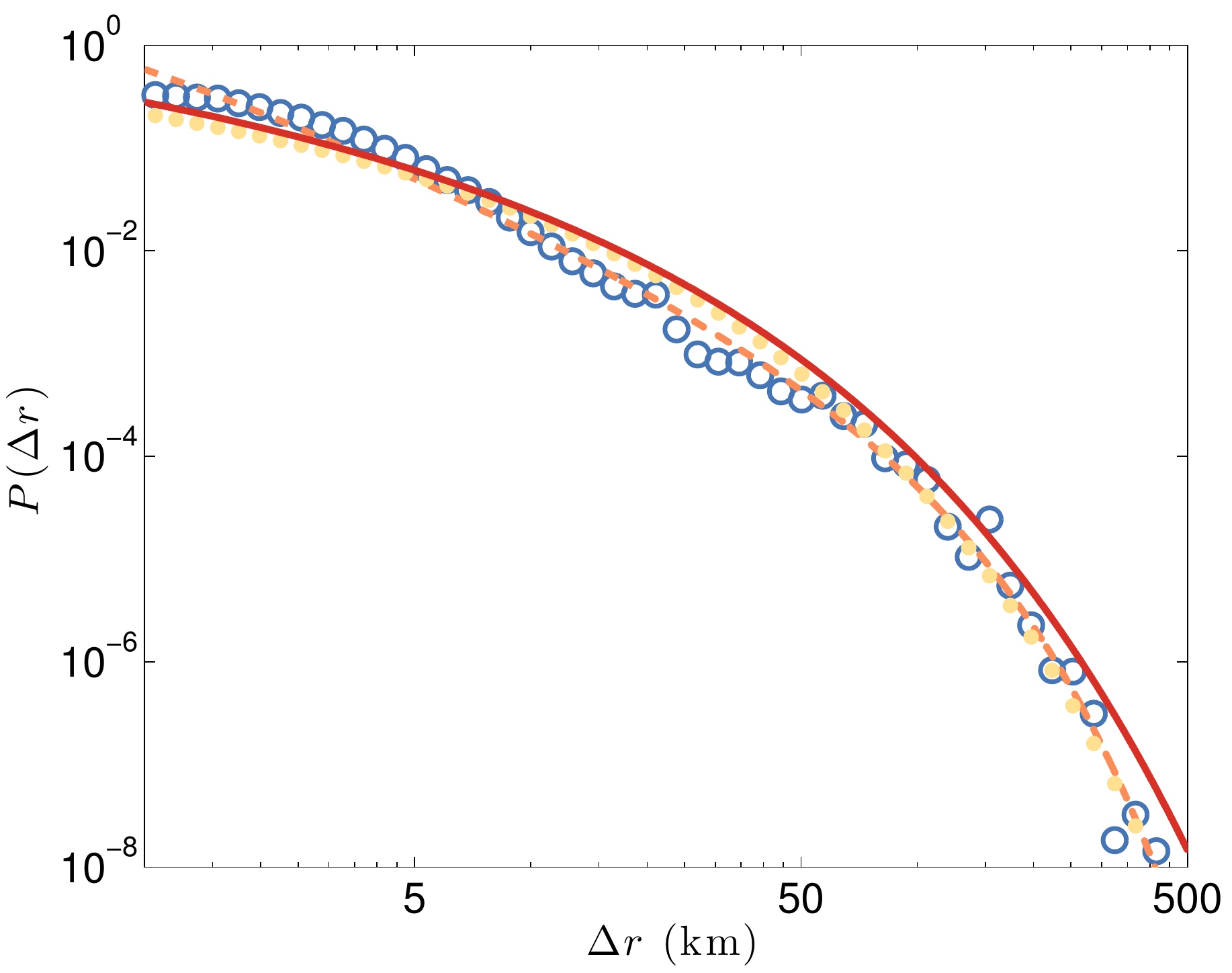} &
\raisebox{2.3cm}{\rotatebox{90}{(Genova)}} \includegraphics[angle=0, width=0.4\textwidth]{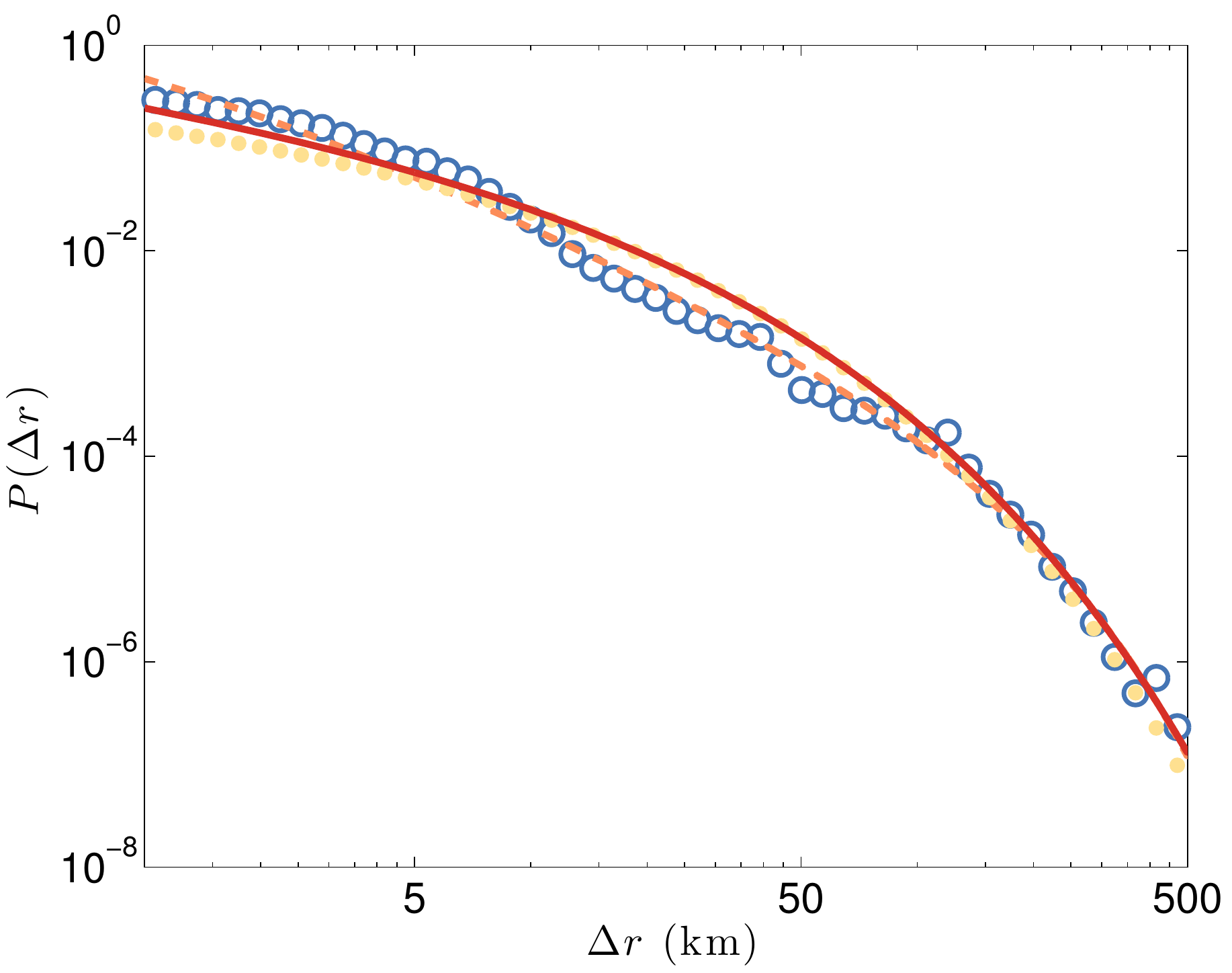} \\
\end{tabular}
\end{center}
\caption{{\bf Displacement distribution for the 6 largest Italian cities.} Similarly to Fig.~\ref{distances} left, we propose here the prediction of our model for the distribution $P(\Delta r)$ in the six largest Italian cities. The blue circles represent the empirical distribution. The orange dash line represents the fit with a Truncated Power Law~\cite{Gonzalez:2008} and the yellow dots with a Stretched Exponential (Eq.~(\ref{randomAccPr}). The red solid line the prediction Eq.~(\ref{eq:dr}). The fit values for the cities are in the ranges $p' = [0.97,1.81]\ \mathrm{jumps}/h$, $\delta v' = [13.2, 18.5]\ \mathrm{km/h}$, $\overline t =  [0.30, 0.37]\ \mathrm{h}$ while $v_0 = 16.5$ km/h.
}
\label{rCities}
\end{figure*}

We note that for single cities there are of course local differences and deviations from the straight line that are probably due to the in-homogeneity of structure of the road network, but as can be seen in Fig.~\ref{vCities} the linear trend is a good fit. Moreover, All cities seem to have a similar average speed of $\approx 16.5 km/h$ at $t = 5$ minutes. We use this value as $v_0$ and estimate, for each of the 6 cities, $p'$ and $\delta v'$ by fitting the conditional probability $P(\overline v|t)$ between $v_0$ and 130 km/h. Finally, we also compute the average travel-time $\overline t$ for each city
This allows us to predict the exact shape of $P(\Delta r)$ for trajectories of drivers living in  different cities. In Fig.~\ref{rCities}, we show that this prediction (red solid line) is  correct, and can be compared with the a-posteriori fit with a Truncated Power Law (orange dash line) or the Stretched Exponential, Eq.~(\ref{randomAccPr}) (yellow dots).


\clearpage

\subsection*{Saddle-point analysis of $P(\Delta r)$.}

The expression for $P(\Delta r)$ can be rewritten
as 
\begin{equation}
P(\Delta r)= \int \frac{\mathrm{d}t}{t \overline t\delta v'} \mathrm{e}^{-F(\Delta r,t)}
\end{equation}
where 
\begin{equation}
F(\Delta r,t)=\left(p'+\frac{1}{\overline t}\right)-\frac{r-v_0t}{\delta v't}\log(p't)+\log\Gamma\left(1+\frac{r-v_0t}{\delta v' t}\right)
\end{equation}
At large $r$ and $t$ this integral is governed by the saddle point defined by $dF/dt=0$ which reads
\begin{equation}
0=p'+\frac{1}{\overline t}-\frac{1}{\delta v'}\left(-\frac{r}{t^2}+\frac{r-v_0t}{t^2}\right)-\frac{r}{\delta v' t^2}\Psi\left(1+\frac{r-v_0t}{\delta v't}\right)
\end{equation}
where $\Psi$ is the Digamma function. We then obtain for large times, the scaling $r\sim t^2$, which implies the following behavior
\begin{equation}
P(\Delta r)\sim \mathrm{e}^{-\Delta r^{1/2}}
\end{equation}
We note that this behavior will also be recovered even if we have Gaussian velocity distribution as it is observed for public transportation.


\begin{thebibliography}{99}

\bibitem{Brockmann:2006}
Brockmann, D., Hufnagel, L. \& Geisel, T.
The scaling laws of human travel. 
{\it Nature} {\bf 439}, 462--465 (2006).

\bibitem{Gonzalez:2008}
Gonz\'alez, M.C., Hidalgo, C.A. \& Barab\'asi, A.-L.
Understanding individual human mobility patterns. 
{\it Nature} {\bf 453}, 779--782 (2008).

\bibitem{Rhee:2011}
Rhee, I., Shin, M., Hong, S., Lee, K. \& Kim, S.
On the levy-walk nature of human mobility.
{\it ACM Transactions on networking} {\bf 19}, 630--643 (2011).

\bibitem{Raichlen:2013}
Raichlen, D. A. et al. 
Evidence of L\'evy walk foraging patterns in human hunter-gatherers
{\it Proc Natl Acad Sci USA} {\bf 111} 728--733 (2013).

\bibitem{Book:foraging}
Viswanathan, G. M., Da Luz, M. G., Raposo, E. P. \& Stanley, H. E. 
{\it The physics of foraging: an introduction to random searches and biological encounters.} Cambridge University Press (2011).

\bibitem{Benhamou:2007}
Benhamou, S.
How many animals really do the L\'evy walk?
{\it Ecology} {\bf 88}, 1962--1969 (2007).

\bibitem{Gallotti:2015ttb}
Gallotti, R., Bazzani, A. \& Rambaldi, S.
Understanding the variability of daily travel-time expenditures using GPS trajectory data.
{\it EPJ Data Science} {\bf 4}, 1--14 (2015).

\bibitem{Song:2010} 
Song, C., Koren, T., Wang, P. \& Barab\'asi, A.-L.
Modelling the scaling properties of human mobility.
{\it Nature Phys} {\bf 6}, 818--823 (2010).


\bibitem{Axhausen:1992}
Axhausen, K.W. \& G\"arling, T.
Activity-based approaches to travel analysis: conceptual frameworks, models, and research problems.
{\it Transport Reviews} {\bf 12}, 323--341 (1992).

\bibitem{Colizza:2007}
Colizza, V., Barrat, A., Barthelemy, M., Valleron, A.J. \& Vespignani, A.
Modeling the worldwide spread of pandemic influenza: baseline case and containment interventions. 
{\it PLoS medicine} {\bf 4}, 95 (2007).

\bibitem{Balcan:2009}
Balcan, D. et al.
Multiscale mobility networks and the spatial spreading of infectious diseases. 
{\it Proc Natl Acad Sci USA} {\bf 106}, 21459--21460 (2009).

\bibitem{Makse:1995}
Makse, H.A., Havlin, S. \& Stanley, H.E.
Modelling urban growth patterns. 
{\it Nature} {\bf 377}, 608--612 (1995).

\bibitem{Bettencourt:2013} 
Bettencourt, L.M., Lobo, J., Helbing, D., Kuhnert, C. \& West, G.B. 
Growth, innovation, scaling, and the pace of life in cities. {\it Proc Natl Acad Sci USA} {\bf 104}, 7301-7306 (2007). 

\bibitem{Louf:2014}
Louf, R. \& Barthelemy, M. 
How congestion shapes cities: from mobility patterns to scaling.
{\it Sci Rep} {\bf 4}, 5561 (2014).


\bibitem{Bazzani:2010}
Bazzani, A., Giorgini, B., Rambaldi, S., Gallotti, R. \& Giovannini, L. 
Statistical laws in urban mobility from microscopic GPS data in the area of Florence.
{\it J Stat Mech} {\bf 2010}, P05001 (2010).

\bibitem{Gallotti:2012}
Gallotti, R., Bazzani, A. \& Rambaldi, S. 
Toward a statistical physics of human mobility.
{\it Int J Mod Phys C} {\bf 23}, 1250061 (2012).

\bibitem{Liang:2012}
Liang, X., Zheng, X., Lv, W., Zhu, T. \& Xu, K. 
The scaling of human mobility by taxis is exponential. 
{\it Physica A} {\bf 391}, 2135 (2012). 

\bibitem{Edwards:2007}
Edwards, A.M. et al. 
Revisiting L\'evy flight search patterns of wandering albatrosses, bumblebees and deer. 
{\it Nature} {\bf 449}, 1044--1048 (2007).

\bibitem{Edwards:2011}
Edwards, A.M.,
Overturning conclusions of L\'evy flight movement patterns by fishing boats and foraging animals.
{\it Ecology} {\bf 92}, 1247--1257 (2011).

\bibitem{Jansen:2012}
Jansen, V.A.A., Mashanova, A. \& Petrovskii, S.
Comment on ``L\'evy  Walks Evolve Through Interaction Between Movement and Environmental Complexity''
{\it Science} {\bf 335}, 918 (2012).

\bibitem{Han:2011}
Han, X.-P., Hao, Q., Wang, B.-H. \& Zhou, T.
Origin of the scaling law in human mobility: Hierarchy of traffic systems.
{\it Phys Rev E} {\bf 83}, 036117 (2011).

\bibitem{Lenormand:2014}
Lenormand, M. et al.
Cross-checking different sources of mobility information. 
{\it PLoS ONE} {\bf 9}, e105184 (2014).

\bibitem{Barabasi:2005}
Barab\'asi, A.-L.
The origin of bursts and heavy tails in human dynamics. 
{\it Nature} {\bf 435}, 207--211 (2005).

\bibitem{Zaburdaev:2008}
Zaburdaev, V., Schmiedeberg, M \& Stark, H.
Random walks with random velocities. 
{\it Phys Rev E} {\bf 78}, 011119 (2008).

\bibitem{Zhao:2015}
Zhao, K., Musolesi, M., Hui, P., Rao, W. \& Tarkoma, S.
Explaining the power-law distribution of human mobility through transportation modality decomposition. 
{\it Sci Rep} {\bf 5}, 9136 (2015).

\bibitem{Roth:2011}
Roth, C., Kang, S.M., Batty, M. \& Barthelemy, M. 
Structure of Urban Movements: Polycentric Activity and Entangled Hierarchical Flows.
{\it PLoS ONE} {\bf 6}, e15923 (2011).

\bibitem{Kolbl:2003}
K\"olbl, R., Helbing, D.
Energy laws in human travel behaviour.
{\it New J Phys} {\bf 5}, 48.1--48.12 (2003).

\bibitem{Gallotti:2013phd}
Gallotti, R.,
{\it Statistical physics and modeling of human mobility},
Ph.D Thesis, University of Bologna, Italy, 79--80 (2013).

\bibitem{Codling:2011}
Codling, E.A. \& Plank, M.J. 
Turn designation, sampling rate and the misidentification of power laws in movement path data using maximum likelihood estimates. 
{\it Theor Ecol} {\bf 4}, 397--406 (2011).


\bibitem{Gallotti:2014}
Gallotti, R. \& Barthelemy, M.
Anatomy and efficiency of urban multimodal mobility.
{\it Sci Rep} {\bf 4}, 6911 (2014).


\bibitem{Bazzani:2011}
Bazzani, A. et al.
Towards congestion detection in transportation networks using GPS data.
{\it IEEE International Conference on Privacy, Security, Risk, and Trust, and IEEE International Conference on Social Computing} (2011).



\bibitem{Vespignani:2012}
Vespignani, A.
Modelling dynamical processes in complex socio-technical systems.
{\it Nature Phys} {\bf 8}, 32--39 (2012).

\bibitem{Axhausen:2002}
Axhausen, K.W., Zimmermann, A., Sch\"onfelder, S., Rindsf\"user, G. \& Haupt, T. 
Observing the rhythms of daily life: A six-week travel diary. 
{\it Transportation} {\bf 29}, 95--124 (2002).

\bibitem{Yan:2013}
Yan, X.-Y., Han, X.-P., Wang, B.-H. \& Zhou, T.
Diversity of individual mobility patterns and emergence of aggregated scaling laws. 
{\it Sci Rep} {\bf 3}, 2678 (2013).

\bibitem{Liang:2013}
Liang, X., Zhao, J., Dong, L. \& Xu, K.
Unraveling the origin of exponential law in intra-urban human mobility.
{\it Sci Rep} {\bf 3}, 2983 (2013).



\bibitem{Kang:2012}
Kang, C., Ma, X., Tong, D. \& Liu, Y.
Intra-urban human mobility patterns: An urban morphology perspective.
{\it Physica A} {\bf 391}, 1702--1717 (2012).

\bibitem{Cheng:2011}
Cheng, Z., Caverlee, J., Lee, K. \& Sui, D.Z. 
Exploring millions of footprints in location sharing services.
{\it ICWSM 2011} 81--88 (2011).

\bibitem{Noulas:2012}
Noulas, A., Scellato, S., Lambiotte, R., Pontil, M. \& Mascolo, C.
A tale of many cities: universal patterns in human urban mobility. 
{\it PLoS ONE} {\bf 7}, e37027 (2012) .

\bibitem{Liu:2014}
Liu, Y., Sui, Z., Kang, C. \& Gao, Y.
Uncovering patterns of inter-urban trip and spatial Interaction from social media check-in data. 
{\it PLoS ONE} {\bf 9}, e86026 (2014).

\bibitem{Hawelka:2014}
Hawelka, B. et al.
Geo-located Twitter as proxy for global mobility patterns. 
{\it Cartogr Geogr Inf Sci} {\bf 41}, 260--271 (2014).

\bibitem{Liu:2012}
Liu, Y., Kang, C., Gao, S., Xiao, Y. \& Tian, Y.
Understanding intra-urban trip patterns from taxi trajectory data.
{\it J geogr syst} {\bf 14.4}, 463--483 (2012).

\bibitem{Wang:2015}
Wang, W., Pan, L., Yuan, N., Zhang, S. \& Liu, D.
A comparative analysis of intra-city human mobility by taxi
{\it Physica A} {\bf 420}, 134--147  (2015).

\bibitem{Liu:2015}
Liu, H., Chen, Y.-H. \& Liha, J.-S.
Crossover from exponential to power-law scaling for human mobility pattern in urban, suburban and rural areas
{\it EPJ B} {\bf 88}, 117 (2015).

\bibitem{Tang:2015}
Tang, J., Liu, F., Wang, Y. \& Wang, H.
Uncovering urban human mobility from large scale taxi GPS data
{\it Physica A} {\bf 438}, 140--153 (2015).



\bibitem{Sagarra:2015}
Sagarra, O., Szell, M., Santi, P., Diaz-Guilera, A. \& Ratti, C. 
Supersampling and network reconstruction of urban mobility.
{\it PLoS ONE} {\bf 10}, e0134508 (2015).

\bibitem{Simini:2012}
Simini, F., Gonz\'alez, M.C., Maritan, A. \& Barab\'asi, A.-L. 
A universal model for mobility and migration patterns.
{\it Nature} {\bf 484},96 (2012).

\bibitem{Louail:2014}
Louail, T. et al. 
From mobile phone data to the spatial structure of cities
{\it Sci Rep} {\bf 4}, 5276 (2014).

\bibitem{Garling:2003}
G\"arling, T. \& Axhausen K.W. 
Introduction: Habitual travel choice.
{\it Transportation} {\bf 30}, 1--11 (2003)

\bibitem{Gallotti:2013}
Gallotti, R., Bazzani, A., Degli Esposti, M. \& Rambaldi, S. 
Entropic measures of individual mobility patterns.
{\it J Stat Mech} {\bf  2013}, P10022 (2013).

\bibitem{Proekt:2012}
Proekt, A., Banavar, J.R., Maritan, A. \& Pfaff, D.W. Scale invariance in the dynamics of spontaneous behavior. {\it Proc Natl Acad Sci of USA} {\bf 109}, 10564--10569 (2010). 

\bibitem{Stouffer:2006}
Stouffer, D.B., Malmgren, R.D. \& Amaral L.A.N. 
Log-normal statistics in e-mail communication patterns. 
arXiv:physics/060527  (2006).

\bibitem{Gallotti:2015mtn}
Gallotti, R. \& Barthelemy, M.
The multilayer temporal network of public transport in Great Britain.
{\it Sci Data} {\bf 2}, 140056 (2015).

\bibitem{Strano:2015}
Strano, E., Shay, S., Dobson, S. \& Barthelemy, M. 
Multiplex networks in metropolitan areas: generic features and local effects. 
{\it J R Soc Interface} {\bf 12}, 20150651 (2015).

\bibitem{Metz:2008}
Metz, D.
The myth of travel time saving.
{\it Transp Rev} {\bf 28}, 321--336  (2008).




\end{thebibliography}
\end{document}